\documentclass[twocolumn]{aastex701}
\usepackage{caption} 
\usepackage{graphicx}
\usepackage{amsmath}
\usepackage{subcaption}
\newcommand{\msun}{{M_\odot}}
\newcommand{\mstar}{M_\star}

\usepackage{soul}

\newcommand{\nautilus}{{\sc nautilus}}
\newcommand{\bagpipes}{{\sc bagpipes}}
\newcommand{\eazy}{{\sc eazy}}

\begin{document}

\title{(Re)solving the Complex Multiscale Morphology and V-shaped Spectral Energy Distribution of a Newly Discovered Strongly Lensed Little Red Dot in A383}

\author[orcid=0009-0005-2295-7246,sname='Josephine Baggen']{Josephine F.W. Baggen}
\affiliation{Department of Astronomy, Yale University, New Haven, CT 06511, USA}
\email[show]{josephine.baggen@yale.edu}
\author[orcid=0000-0002-8282-9888,sname='Pieter van Dokkum']{Pieter van Dokkum}
\affiliation{Department of Astronomy, Yale University, New Haven, CT 06511, USA}
\email{pieter.vandokkum@yale.edu}
\author[orcid=0000-0002-2057-5376,sname='Ivo Labb\'e']{Ivo Labb\'e}
\affiliation{Centre for Astrophysics and Supercomputing, Swinburne University of Technology, Melbourne, VIC 3122, Australia}
\email{ivolabbe@gmail.com}
\author[orcid=0000-0003-2680-005X,sname='Gabriel Brammer']{Gabriel Brammer}
\affiliation{Cosmic Dawn Center (DAWN), Niels Bohr Institute, University of Copenhagen, Jagtvej 128, K{\o}benhavn N, DK-2200, Denmark}
\email{gabriel.brammer@nbi.ku.dk}

\begin{abstract}
We present a luminous "little red dot" (LRD) at $z$ = 6.027, doubly imaged by the galaxy cluster A383 and observed with the James Webb Space Telescope (JWST) NIRCam. Owing to its large magnifications, $\mu$ $\sim$ 11 for image A383-LRD1A and $\mu$ $\sim$ 7 for A383-LRD1B, the system is exceptionally bright and highly stretched, providing a rare, spatially resolved view of an LRD. The images reveal a complex morphology with a compact red dot, a spatially offset blue dot, and faint emission bridging and surrounding the two. After correcting for lensing, the blue and red dots have rest-frame UV and optical sizes of $\sim 60$ pc and $\lesssim 150$ pc, respectively, while extended emission traced most clearly in F356W ([O\,III]+H$\beta$) reaches scales of order $\sim 1$ kpc.
Spatially resolved spectral energy distribution (SED) analysis reveals that the characteristic V-shaped SED arises from the superposition of a flat UV continuum from the blue dot, consistent with a young stellar population, and a steep red SED from the red dot, which, based on Hubble Space Telescope+JWST photometry alone, admits a straightforward stellar interpretation as a massive, heavily dust-attenuated component. However, this interpretation is challenged by Atacama Large Millimeter/submillimeter Array dust continuum upper limits, suggesting a nonstellar origin such as dense gas configurations. 
Separated by only $\sim$ 300 pc in the source plane, these components would blend into a single compact source in unlensed observations with the canonical LRD colors. 
This system therefore provides a rare opportunity to resolve the internal structure of an LRD and to begin unraveling the physical nature of this population.
\end{abstract}



\section{Introduction} 

The James Webb Space Telescope (JWST) is transforming our view of the earliest phases of galaxy and black hole formation. Among its most surprising discoveries is a population of compact red sources seen across many surveys \citep[e.g.,][]{Harikane2023,Kocevski2023, Labbe2023, Kokorev2024, Maiolino2024_jades, Matthee2024, Akins2024}, now colloquially termed “little red dots” (LRDs).

The properties of LRDs are unusual, making them difficult to interpret within standard models.
In the optical, they show extremely red continua \citep{Furtak2024}, compact morphologies \citep[e.g.,][]{Akins2023, Baggen2023, Baggen2024}, broad Balmer lines \citep{Kocevski2023,Harikane2023, Greene2024, Maiolino2024_jades,Matthee2024}, and frequent Balmer breaks \citep{Labbe2024,Wang2024_balmer, Graaff2025_Cliff, Naidu2025_BHstar}. 
At the same time, LRDs exhibit blue colors in the rest-frame UV \citep[e.g.,][and references therein]{Kocevski2025ApJ...986..126K}. Short-wavelength JWST imaging further reveals that this UV emission is often irregular and spatially offset from the red emission \citep{Baggen2024, Chen2025,Rinaldi2025ApJ...992...71R_notdot, Baggen_2026, Ren2026ApJ..1003..150R}.
When combined, the integrated light from the rest-UV to rest-optical forms a distinctive “V-shaped” spectral energy distribution (SED), with the inflection point occurring consistently at the Balmer limit \citep{Kocevski2025ApJ...986..126K, Hviding2025,Setton2025ApJ...995..118S_vshape}.

 A leading model to explain these features is that they are produced by active galactic nuclei \citep[AGNs;][]{Greene2024, Matthee2024, Kocevski2025ApJ...986..126K}. However, LRDs lack nearly all canonical signatures of accretion: They are X-ray faint \citep{Ananna2024}, radio-quiet \citep{Gloudemans2025}, lack hot dust emission \citep{PerezGonzalez2024A, Williams2024, Setton2025_dust}, and lack hard ionizing photons \citep{Wang2025_ionizing}, 
 show little or no variability \citep[e.g.,][]{Burke2025},
 and their SEDs favor stellar over AGN templates \citep{PerezGonzalez2024A, Carranza-Escudero2025}. 
 This has motivated more exotic scenarios in which the black hole is hidden within dense, optically thick gas \citep{Juodzbalis2024,Graaff2025_Cliff,Graaff2025_unified,InayoshiMaiolino2025,Ji2025,Kido2025MNRAS.544.3407K, Liu2025ApJ...994..113L, Naidu2025_BHstar,
 Asada2026arXiv260110573A,
 DEugenio2025,
 Inayoshi2026ApJ..1000...90I, 
 Kokorev2025arXiv251107515K,
 Liu2026arXiv260302317L,
 MadauMaiolino2026arXiv260222386M, 
 Matthee2026arXiv260317667M,
 Pacucci2026arXiv260114368P,
 Rusakov2026Natur.649..574R,
Sneppen2026arXiv260118864S,  Sun2026arXiv260120929S}, 
 producing an A-type star-like spectrum with a strong Balmer break. Such “black-hole–star systems” \citep[BH*;][]{Naidu2025_BHstar, Graaff2025_Cliff} may lack a stellar host entirely \citep{Juodvzbalis2025_BH_measurement}. A related possibility is that some LRDs are late-stage quasi-stars \citep{Begelman2006, Begelman2026}; synthetic spectra from these models show promising agreement with observations \citep{Santarelli2025}. In all these scenarios, the UV emission is thought to arise from regions close to the black hole, either in the form of scattered AGN light \citep{Furtak2023_triple,Greene2024,Leung2025} or nebular continuum from low-density gas \citep{Chen2025}.

An alternative idea is that the light in both the rest-frame UV and optical comes from stars. In the optical, this would imply both evolved stellar populations and significant dust absorption
\citep[e.g.,][]{Akins2023,Labbe2023, Barro2024,Wang2024_balmer}, yielding extreme stellar masses \citep{Labbe2023,Wang2024_balmer}. The high
masses combined with the small sizes
of LRDs lead to extremely high inferred stellar densities,
and the observed broad Balmer lines are explained by virial motions alone \citep{Baggen2024}. 
In this scenario, the UV emission simply traces regions
of  relatively unobscured star formation,
consistent with the absence of Mg\textsc{ii}~$\lambda2800$ \citep{Akins2025} expected for the AGN scattering scenario, spatially resolved narrow lines and UV continua \citep{Killi2024}, and the extended, complex, and sometimes offset UV morphologies \citep{Baggen2024, Killi2024, Chen2025, Rinaldi2025ApJ...992...71R_notdot, Baggen_2026}.

Much of the effort to discriminate between these scenarios
has been focused on the interpretation of the spectra and SEDs
of LRDs. The internal structure of LRDs provides -- in principle -- an important additional constraint. In the ``naked'' black hole scenarios, all emission is centered on the black hole and is extremely compact; as an example, the "quasi-star" picture should have a size of only up to a few thousand au \citep[0.02 pc;][]{Begelman2026}.
By contrast, in stellar-dominated models, the galaxies should show a spatially resolved morphology at sufficiently high spatial resolution, with the size of the dominant mass component given by
$r_e \sim M/\sigma^2$.

The only practical way to obtain the necessary resolution is through
strong gravitational lensing.
To date, only a single LRD has been studied in such detail: a triply lensed, intrinsically faint system in A2744 (A2744-QSO1) magnified by a factor of $\sim7$ \citep{Furtak2023_triple, Furtak2024,Greene2024}. 
This source has been observed with integral field unit (IFU) spectroscopy at high spatial resolution \citep{Ji2025, Deugenio2025_lrd_furtak} and supports a minimal host-galaxy contribution to the observed emission \citep{Juodvzbalis2025_BH_measurement}.
Here we present a second strongly lensed LRD that is more than an order of magnitude brighter than A2744-QSO1. 
We also highlight recent independent analyses of this system by \citet{Knudsen2025} and \citet{Golubchik2025}.

Throughout this paper, we adopt the flat Planck18 CDM cosmology \citep[see Table 2 in][]{Planck2020}, and report all magnitudes in the AB system.

\begin{figure*}[htp!]
    \centering
    \includegraphics[width=0.95\linewidth]{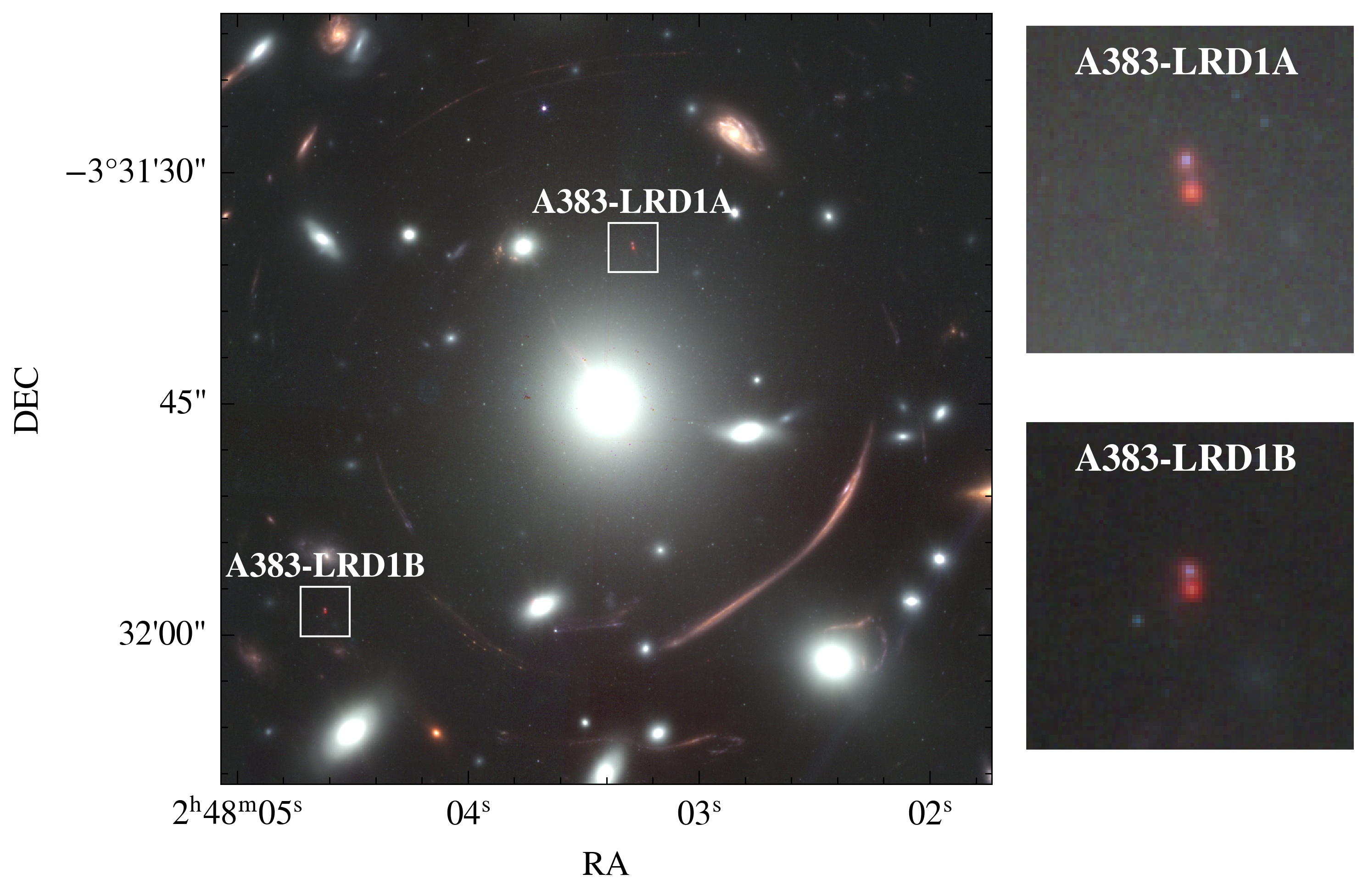}
    \vspace{0.3cm}
    \includegraphics[width=\linewidth, trim={0cm, 0.0cm, 0cm 0cm}, clip]{cutouts_final_1.0_arcsec_0.png}
    \includegraphics[width=\linewidth, trim={0cm, 0cm, 0cm 1.5 cm}, clip]{cutouts_final_1.0_arcsec_1.png}
    \caption{Top: red, green, and blue (RGB) composite image of A383, constructed using F090W + F115W + F150W for blue, F200W + F210M for green, and F277W + F356W + F444W for red. Inset cutouts show the doubly lensed system (A383-LRD1A and A383-LRD1B) and are $1\farcs6 \times 1\farcs6 $. Bottom: multiband cutouts ($1\farcs0 \times 1\farcs0 $) for the images across HST and JWST filters. 
    }
    \label{fig:rgb}
\end{figure*}

\begin{figure*}[htp!]
   \centering
\begin{tabular}{lcccccccccc}
\hline
ID & RA & DEC & F435W & F606W & F814W & F125W & F140W & F160W & F090W & F115W \\
\hline
A383-LRD1A & 42.01369 & -3.52636 & -33$\pm$33 & -5$\pm$22 & 174$\pm$49 & 477$\pm$74 & 390$\pm$80 & 436$\pm$95 & 312$\pm$24 & 442$\pm$32 \\
A383-LRD1B & 42.01924 & -3.53292 & 7$\pm$38 & -25$\pm$21 & 75$\pm$26 & 257$\pm$44 & 254$\pm$45 & 252$\pm$51 & 243$\pm$21 & 322$\pm$28 \\
\hline
\multicolumn{3}{c}{} & F150W & F200W & F210M & F277W & F300M & F356W & F410M & F444W \\
\hline
\multicolumn{3}{c}{} & 407$\pm$35 & 529$\pm$40 & 461$\pm$38 & 672$\pm$71 & 707$\pm$61 & 1931$\pm$49 & 1071$\pm$40 & 2237$\pm$40 \\
\multicolumn{3}{c}{} & 289$\pm$28 & 350$\pm$33 & 303$\pm$33 & 430$\pm$30 & 461$\pm$32 & 1289$\pm$30 & 661$\pm$28 & 1393$\pm$25 \\
\hline
\end{tabular}
\captionof{table}{Aperture fluxes (in nanojanskys, observed) for A383-LRD1A
and A383-LRD1B 
across HST/JWST filters.
Coordinates correspond to the F444W centroid positions of each image. 
\label{tab:photometry}
    }
     \vspace{1em}  
    \includegraphics[width=0.6\linewidth]{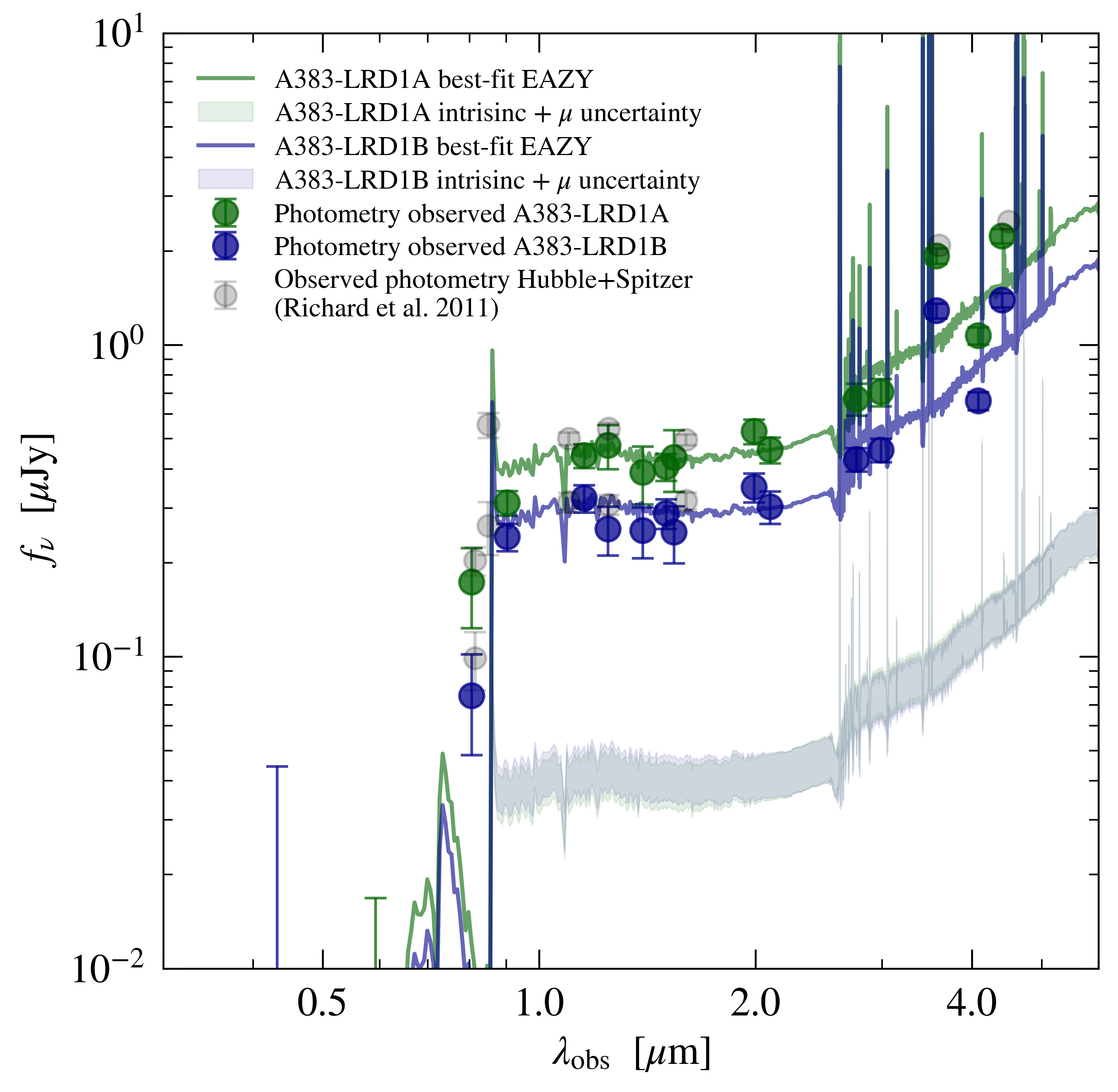}
    \captionof{figure}{Spectral energy distribution (SED) derived from our aperture photometry for A383-LRD1A (green points) and A383-LRD1B (blue points). The solid lines show the best-fit model to the total SED obtained with \eazy. The SED exhibits the characteristic “double-break” (or “V-shape”), corresponding to the Lyman and Balmer breaks, and its UV to optical colors are typical of the LRD population.
    Adopting the reported magnifications of $\mu_A = 11.4\pm1.9$ and $\mu_B = 7.3\pm1.2$ by \citet{Richard2011}, we compute the intrinsic SED, with the uncertainties in $\mu$ represented by the shaded regions. The observed flux ratios (A383-LRD1A/A383-LRD1B) are in perfect agreement with the predicted $\mu_A/\mu_B$. 
    }
    \label{fig:SED}
\end{figure*}

\section{A Strongly Lensed LRD Behind A383}
\subsection{Observations and Identification}
\label{sec:imaging}
The source was identified in NIRCam imaging of the cluster A383.
The primary data used in this work are JWST/NIRCam observations from the VENUS program (Program ID: GO 6882; PI: S. Fujimoto), supplemented by Hubble Space Telescope (HST) Advanced Camera for Surveys (ACS) and Wide Field Camera 3 (WFC3) imaging from the CLASH survey \citep{Postman2012}. All data were reduced using the \texttt{grizli} pipeline v7.2 \citep{brammer_gabriel_2022_6672538}, with calibration reference files from CRDS context \texttt{jwst\_1293.pmap}. This version includes snowball masking for NIRCam and NIRISS exposures, together with updated bad-pixel masks. The reductions follow the methodology outlined in the PANORAMIC field notebook,\footnote{\url{https://github.com/gbrammer/panoramic-jwst/blob/main/Notebooks/step3-panoramic-mosaics.ipynb}} in which individual exposures are aligned onto a common astrometric frame and combined into mosaics using drizzle resampling.
The final mosaics used in this work contain 16 filters: HST/ACS (F435W, F606W, F814W), HST/WFC3 (F125W, F140W, F160W), and JWST/NIRCam (F090W, F115W, F150W, F200W, F210M, F277W, F300M, F356W, F410M, F444W). 
The NIRCam mosaics have pixel scales of $0\farcs02$ and $0\farcs04$ for the short-wavelength (SW; $\lambda < 2.4\,\mu$m) and long-wavelength (LW; $\lambda > 2.4\,\mu$m) channels, respectively. All HST mosaics are drizzled to $0\farcs04$ pixel$^{-1}$.

We show a composite color image of A383 made from the JWST/NIRCam mosaics in Figure~\ref{fig:rgb}. Two compact red objects are apparent in the cluster field: A383-LRD1A (R.A.=42.01369, decl.=-3.52636) and A383-LRD1B (R.A.=42.01924, decl.=-3.53292).
These sources were first identified as lensed high-redshift galaxies in the CLASH survey, 
with $z_{\rm spec}=6.027$ from the Ly$\alpha$ emission line using DEIMOS spectroscopy. Best-fitting lensing models indicate extreme magnifications of $\mu_A = 11.4\pm1.9$ and $\mu_B = 7.3\pm1.2$ \citep{Richard2011, Stark2015}.

\subsection{Total Photometry and Spectral Energy Distribution}
\label{sec:totalphotometry}
We perform point-spread function (PSF)-matched aperture photometry on $10\arcsec\times10\arcsec$ cutouts centered on both A383-LRD1A and A383-LRD1B using all available HST and JWST imaging. To account for the varying PSFs across instruments and wavelengths, all images are convolved to match the F444W PSF using empirical PSF models and matching kernels from \citet{Weaver2024}.\footnote{We note that the empirical PSFs used for photometry are constructed from the A2744 field. However, for aperture photometry with our adopted aperture of $0\farcs5$, deviations between the PSF model and observed stellar profiles are expected to be minimal, $<1\%$ at these radii \citep{Williams2025ApJ...979..140W}. For structural modeling, we adopt a more detailed treatment of the PSF, as described in Section~\ref{sec:morphfitting}.}
Background subtraction is applied to the PSF-matched images using a sigma-clipped 2D median filter computed in $25\times25$ pixel boxes, followed by an additional $5\times5$ pixel median smoothing to generate a background model.
Photometry is then performed with \texttt{photutils} \citep{larry_bradley_2025_14889440} on the background-subtracted images, using circular apertures with a radius of $0\farcs5$.
Flux uncertainties are estimated empirically by measuring the rms of 1000 empty apertures of the same radius placed in source-free background regions of these images.
Aperture corrections, derived from the PSF-matched models to account for flux outside the measurement aperture, are applied to both the measured fluxes and their uncertainties. 

\begin{figure*}[htp!]
    \centering
    \includegraphics[width=0.45\linewidth, trim={0cm 0cm 0cm 0cm}, clip]{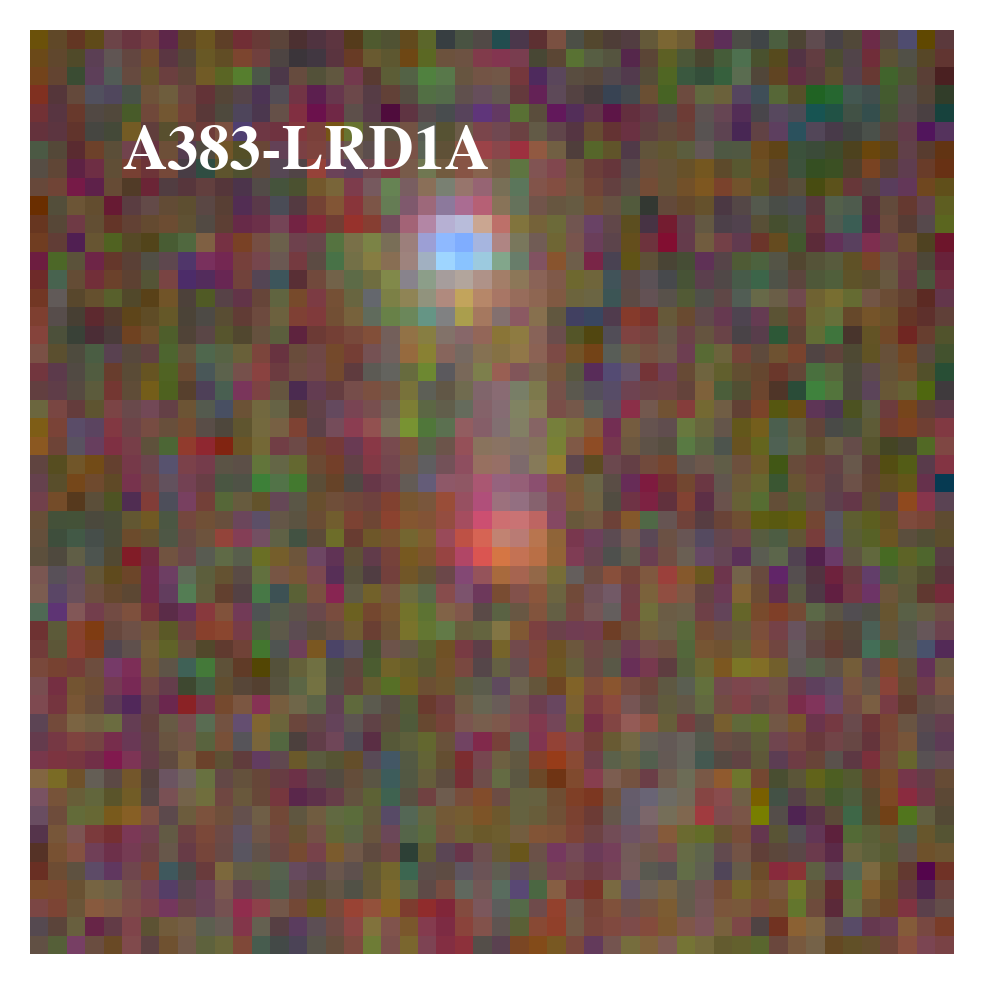}
     \includegraphics[width=0.45\linewidth, trim={0cm 0cm 0cm 0cm}, clip]{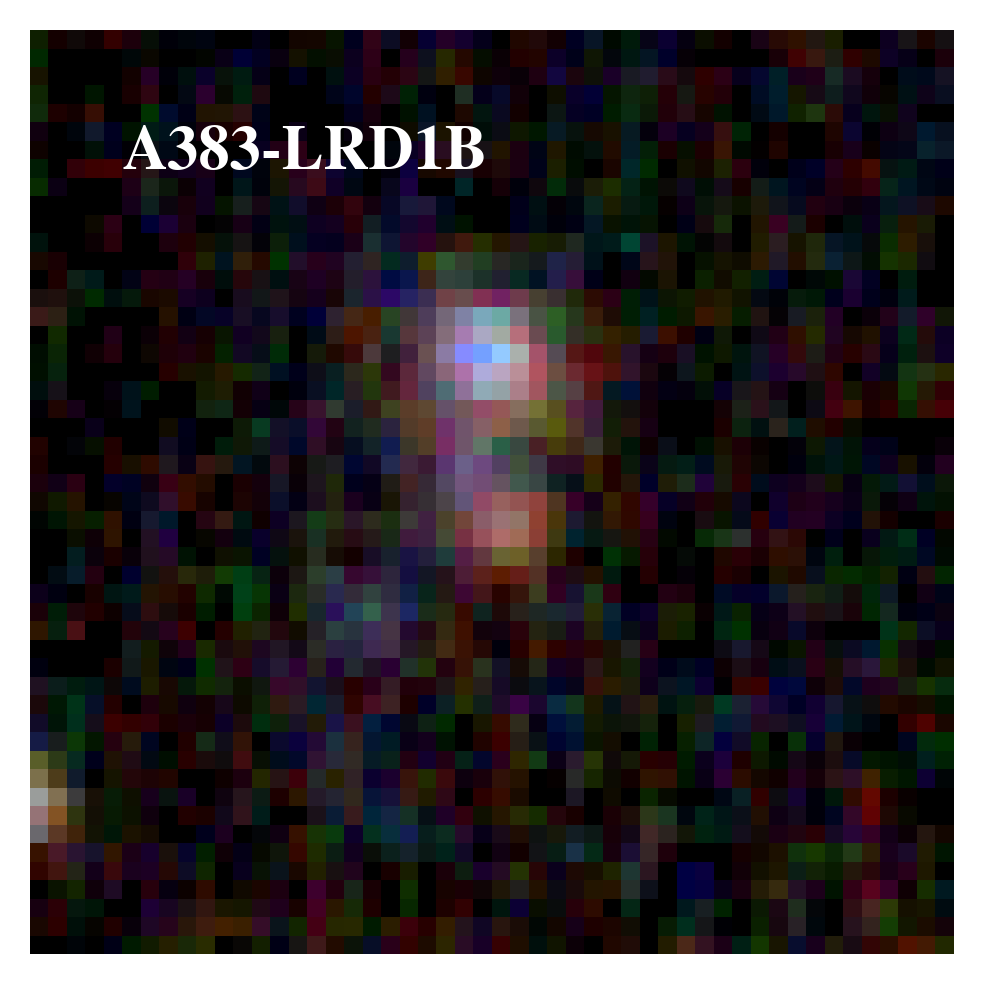}
    \caption{RGB image stamps constructed using filters F200W, F150W, and F115W for the two lensed images: A383-LRD1A (left) and A383-LRD1B (right). Both images are $1 \arcsec\times 1\arcsec$. In this work, we focus on image A383-LRD1B, as it lies in a relatively dark region of the sky with minimal contamination from intracluster light (ICL), whereas A383-LRD1A is embedded in a brighter ICL background. We identify a prominent blue component, red component, and a faint structure bridging and surrounding the two, and model their surface brightness profiles as described in the text.  
    }
    \label{fig:morphology}
\end{figure*}

The resulting photometry (in nanojanskys) is presented in Table~\ref{tab:photometry} and the SED is shown in Figure~\ref{fig:SED}. Our measurements are consistent with previous measurements by \citet{Richard2011}. 
Its red colors, already identified in the Spitzer/IRAC 3.6 and 4.5~$\mu$m bands,
had previously been interpreted as evidence of a strong Balmer break and thus a massive galaxy with an old stellar population \citep{Richard2011}. 
Our new JWST/NIRCam photometry confirms that the discontinuity indeed falls at the Balmer break, producing the characteristic "double-break" (or "V-shape" when plotted in $f_{\lambda}$) SED. The observed colors (for A383-LRD1B) are
\begin{align*}
    \mathrm{(F150W - F200W)} &= 0.2 \\
    \mathrm{(F277W - F356W)} &= 1.2 \\
    \mathrm{(F277W - F444W)} &= 1.3. 
\end{align*}
These colors match the photometric selection criteria now used for selecting LRDs in the literature \citep{Greene2024, Kocevski2025ApJ...986..126K}. We further confirm the source compactness following \citet{Greene2024} using the ratio of F444W flux measured within $0\farcs4$ and $0\farcs2$ apertures (compact = $f_{\rm f444w}(0\farcs4)/f_{\rm f444w}(0\farcs2)<1.5$). 
In the source plane (after demagnification), the system becomes intrinsically more compact (see Section~\ref{sec:lensmodeling_reconstruction}), so this ratio is expected to be even smaller.

This LRD is exceptionally bright, ranking among the brightest known on the sky. The observed fluxes are comparable to the ultraluminous LRD reported by \citet{Labbe2024} and nearly 2 magnitudes brighter than the only other strongly lensed case known to date \citep{Furtak2023_triple, Furtak2024}.

Fitting the total observed photometry with \eazy\, \citep{Brammer2008} (assuming a standard FSPS-based template set; tweak\_fsps\_QSF\_12\_v3) 
yields the solid model curves shown in Figure~\ref{fig:SED}, which reproduce the measured fluxes well. The high fluxes in the F356W and F444W bands are interpreted as strong emission lines (e.g., H$\beta$ + [O\,III] and H$\alpha$), falling within these filters at this redshift.
To illustrate the intrinsic SED, we scale the best-fit model by the magnifications derived by \citet{Richard2011}, with the corresponding uncertainties in $\mu$ represented by the shaded regions. We perform this for both A383-LRD1A and A383-LRD1B individually. Within the uncertainties, the two demagnified SEDs lie essentially on top of each other, lending confidence to the adopted magnification ratios.
The intrinsic best-fit model yields a stellar mass of $\log \mstar/\msun \approx9.7$,
fully consistent with the conclusions of \citet{Richard2011}.
We perform an independent SED fit using \bagpipes\, \citep{Carnall2018}, which uses nested sampling \citep[via \nautilus;][]{Lange2023MNRAS.525.3181L_nautilus}.
We assume an exponentially declining star formation history with broad priors on age (0.005-1 Gyr), $\tau$ (0.01-10 Gyr), and stellar mass ($\log M_{\mathrm{formed}}/\mathrm{M}\odot$ between 6 and 12). We adopt a log-uniform metallicity prior over $10^{-4}$ to $2.5\,Z\odot$, include nebular emission with $\log U$ from -4 to -1, and model dust attenuation using a Calzetti law with $A_V$ between 0 and 6. 
This setup gives consistent properties to those inferred from \eazy.

The interpretation of an old stellar population proposed more than a decade ago, before LRDs were recognized as a distinct class, now sits at the center of current debates about the origin of LRD emission. 
In Section~\ref{sec:sedcomponents} we return to this and reevaluate the SED using the full leverage of the new JWST data, demonstrating that the observed total SED is a superposition of multiple physical components, which must be separated for a physically meaningful interpretation.

%


\section{Component decomposition}
\subsection{Visual Morphology}
The morphology of the lensed galaxy, seen in both images A383-LRD1A and A383-LRD1B, is complex and reveals multiple distinct components, as highlighted in Figure~\ref{fig:morphology}.
The system clearly consists of two bright, spatially distinct components: a compact region that dominates in the rest-frame UV and a compact component that becomes prominent at longer wavelengths.
We further identify a fainter structure between and surrounding these two components.

\subsection{Surface Brightness Profile Fitting}
\label{sec:morphfitting}
We first focus on the morphology of the two brightest components: the compact red source and the neighboring blue component.
The analysis focuses on image A383-LRD1B, as it lies
in a relatively dark region of the sky with minimal contamination from intracluster light (ICL), whereas A383-LRD1A is
embedded in a brighter ICL background, making it less
suitable for reliable profile fitting. 

We perform surface brightness profile fitting using GALFIT \citep{Peng2002, Peng2010x} across all available bands to quantify the morphology. 
In this work, we consider a range of structural models, including S\'ersic profiles \citep{Sersic1968} and PSF models for each component (Appendix~\ref{app:galfit_tests}). As motivated below, we 
adopt a fiducial model in which the blue component is fit with a 
S\'ersic profile and the red component with a PSF.
For the S\'ersic fits, the free parameters include the position ($x,y$), effective radius ($r_{\rm e}$), S\'ersic index ($n$), total magnitude, axis ratio ($b/a$), and position angle. We let $r_{\rm e}$ vary between 0.5 and 50 pixels, $b/a$ between 0.1 and 1, and the magnitude between 1 and 100 mag. Radii are circularized with $r_{\rm e,circ} = \sqrt{b/a}\,r_{\rm e}$. For the PSF fits, only the position and total magnitude are free parameters.

Accurate PSF characterization is essential for the structural modeling, but the exact PSF is not directly known. Theoretical PSFs generated with STPSF \citep[formerly WebbPSF;][]{Perrin2014} have been reported to be sharper than observed stellar profiles \citep[e.g.,][]{Ito_2024}, although recent versions show improved agreement \citep[e.g.,][]{Morishita2024ApJ...963....9M, Weibel2024}. Ideally, one would construct an empirical PSF directly from the science images. However, the A383 field contains only a very limited number of 
suitable stars (only two in the SW bands), making it difficult to derive a reliable empirical PSF \citep[see also][who report similar limitations]{Williams2025ApJ...979..140W, Cloonan2026arXiv260324700C}. 
We therefore follow \citet{Ito_2024} and construct a smoothed version of the STPSF, by smoothing with a Moffat profile, calibrated to match the observed stellar profiles in A383. We adopt the smoothed STPSF as our fiducial PSF in the main analysis.

We explore a suite of structural models to determine how robust the 
decomposition is to model assumptions, including S\'ersic+S\'ersic 
and PSF+PSF configurations (for the red and blue components, respectively), as well as variations in the adopted PSF 
(Appendix~\ref{app:galfit_tests}).
We find that the size of the blue component and the separation between the two components are generally stable across these models, whereas the properties of the red component reveal a significant degeneracy with the surrounding low–surface-brightness emission,  particularly in the SW bands and in F277W/F356W. Motivated by this degeneracy and the extremely compact visual appearance, we adopt a simple fiducial model in which the red component is represented by a PSF and the blue component by a S\'ersic profile, while the remaining extended emission is treated separately (Section~\ref{sec:extendedcomp}).

Prior to fitting, we identify and mask contaminating sources using sigma-clipped statistics with a 5 pixel filter size to estimate the background and rms ($\sigma$). After background subtraction, images are convolved with a 2D Gaussian kernel (FWHM = 3 pixels), and sources above $1\sigma$ are masked. The mask is then refined to preserve both components of the target source while excluding nearby contaminants. The fits are done first in the F200W image, which provides the highest S/N at optimal resolution, in order to determine accurate initial centroid positions for the two components. These F200W-derived positions are then used as initial guesses in the fits to the other bands, and we allow the centroids in both $x$ and $y$ to vary only within $\pm0\farcs04$, corresponding to $\pm$2 pixels in the SW images and $\pm$1 pixel in the LW images.
To explore the wavelength-dependent morphology, we allow the structural parameters to vary independently in each band.

\begin{figure*}[htp!]
    \centering
    \includegraphics[width=\linewidth]{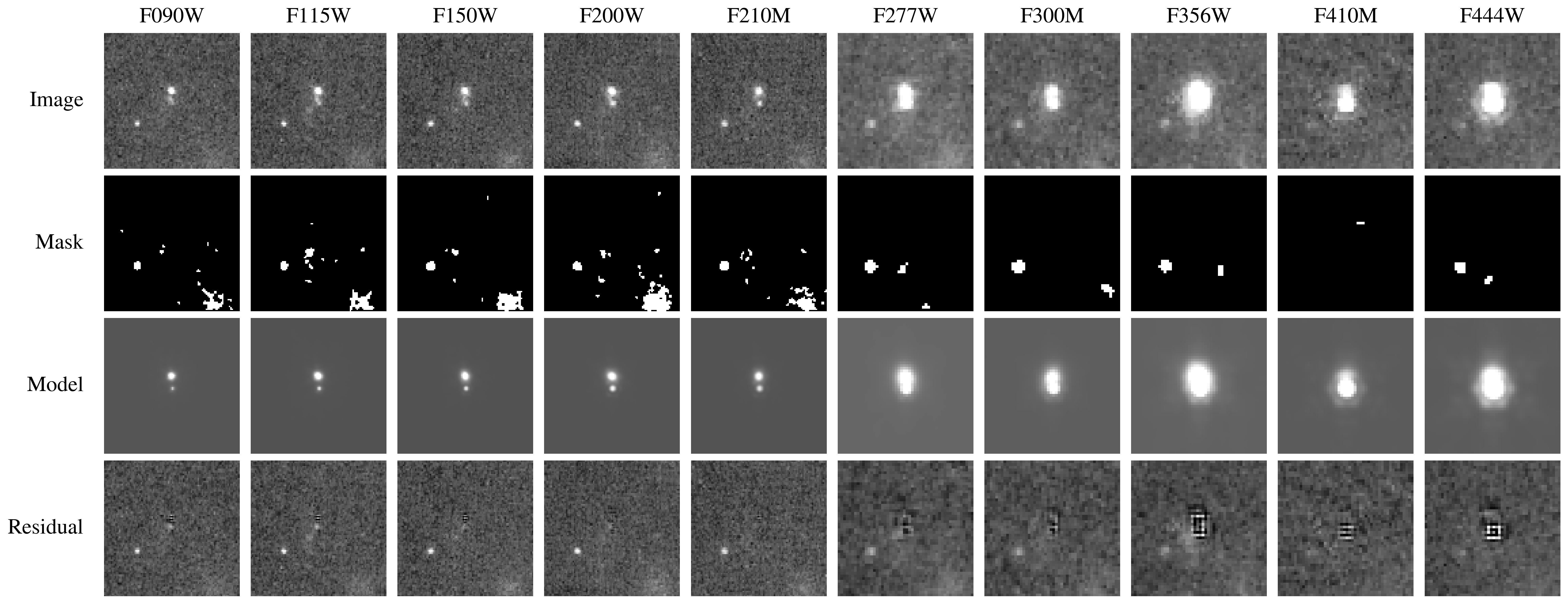}
    \includegraphics[width=0.8\linewidth]{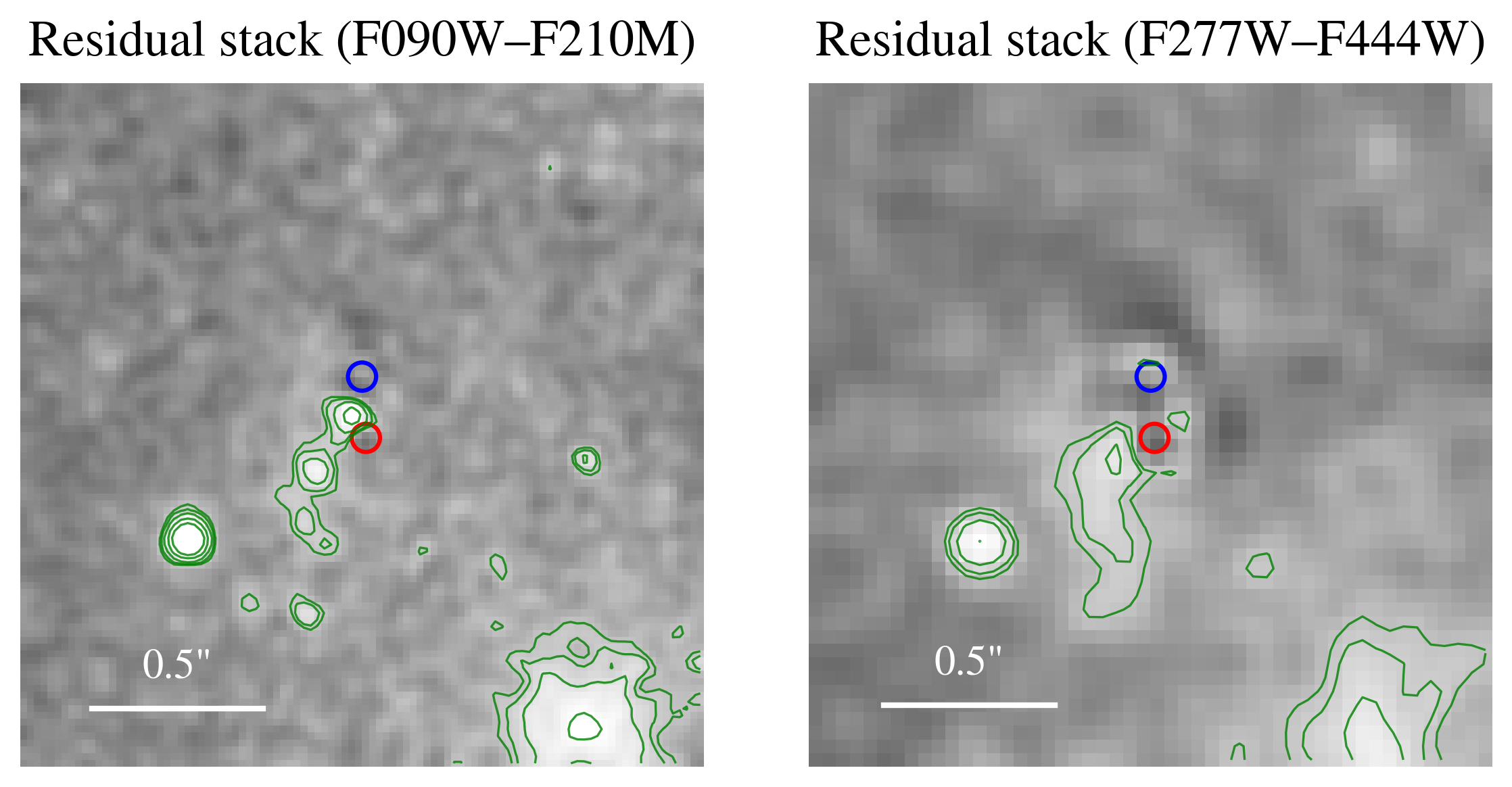}
    \caption{\textit{Top:} GALFIT best-fit models for A383-LRD1B. For each filter, we show (from top to bottom): the original image, the fitting mask, best-fit model, and residual. The overall fits are good, with minimal residuals in most bands. \textit{Bottom:} stacked residual images for the short-wavelength (SW) and long-wavelength (LW) filters. In both stacks, extended emission (green) is clearly detected with a similar spatial extent. In particular, the SW stack reveals bridgelike emission connecting the red and blue components (indicated by open circles for visualization).} 
    \label{fig:galfit_final}
\end{figure*}

We estimate uncertainties on the best-fit parameters using injection–recovery simulations. The best-fit model for each component is inserted into 100 randomly selected empty-sky regions and
convolved with either STPSF or the smoothed STPSF.
For each parameter, we report the best-fit value measured in the real image, while the uncertainties correspond to the 16th–84th percentile range of the recovered
distributions, computed relative to the simulation median. This procedure captures the impact of sky background variations and PSF uncertainties on the recovery of a given model, but  does not account for systematic uncertainties associated with different plausible models or PSF choices, which affect the best-fit values themselves
(Appendix~\ref{app:galfit_tests}).

\subsection{Wavelength-Dependent Structural Parameters}
\label{sec:morphparameters}
The structural parameters derived from our fiducial model are 
presented in Table~\ref{tab:characteristic_scales}.  These values are reported in the observed frame (first column) and do not yet account for gravitational lensing. In Section~\ref{sec:lensmodeling_reconstruction} we reconstruct the intrinsic (source-plane) values (second column).

Figure~\ref{fig:galfit_final} (top panel) shows the fitting results for our fiducial model, in which the blue component is modeled with a S\'ersic profile and the red component with a PSF, using the smoothed STPSF.
For each filter (column), the figure displays the observed image (top), the mask with pixels that are ignored during fitting (second row), the best-fit model (third row), and the residual image (bottom). The GALFIT models match the data closely, with minimal residuals in most bands, and a consistent separation between the blue and red components of $d_{\rm sep}\sim0\farcs18-0\farcs2$ in A383-LRD1B. For the other lensed image (A383-LRD1A), we find a robust separation between the centroid positions of the blue and red components of $d_{\rm sep}\sim0\farcs3$ across the bands. We revisit the comparison between the separations for A383-LRD1A and A383-LRD1B in Section~\ref{sec:lensmodeling_reconstruction}.

\begin{table}[htp!]
\centering
\begin{tabular}{lcc}
\hline
Component & Observed & Intrinsic \\
\hline
Blue comp. ($r_{\rm e,F090W}$) & $\leq0\farcs015$ & $\leq32\pm3\,\mathrm{pc}$\\
Blue comp. ($r_{\rm e,circ,F150W}$) & $0.024^{+0.011}_{-0.003}\arcsec$ & $52^{+25}_{-8}\,\mathrm{pc}$\\
Blue comp. ($r_{\rm e,circ,F200W}$) & $0.034^{+0.006}_{-0.007}\arcsec$ & $73^{+15}_{-16}\,\mathrm{pc}$\\
Blue comp. ($r_{\rm e,circ,F210M}$) & $0.030^{+0.011}_{-0.005}\arcsec$ & $64^{+25}_{-12}\,\mathrm{pc}$\\
Blue comp. ($r_{\rm e,circ,F356W}$) & $0.054^{+0.035}_{-0.004}\arcsec$ & $115^{+76}_{-13}\,\mathrm{pc}$\\
Red comp. ($r_{\rm e,F444W}$) & $\leq0\farcs070$ & $\leq151\pm12\,\mathrm{pc}$\\
Extended comp. ($d_{\rm extend}$) & $\sim0\farcs3-0\farcs5$ & $\sim0.6-1.1\,\mathrm{kpc}$\\
Separation ($d_{\rm sep}$) & $0\farcs18-0\farcs20$ & $300\pm30\,\mathrm{pc}$\\
\hline
\end{tabular}
\caption{
Characteristic spatial scales of the system, measured for image A383-LRD1B. Central 
values correspond to the fiducial model (Section~\ref{sec:morphfitting}). Uncertainties on the observed sizes come from random errors from 
injection-recovery simulations.
Intrinsic physical values are derived using the lensing corrections described in Section~\ref{sec:lensmodeling_reconstruction}. Uncertainties 
on the intrinsic sizes additionally include errors on 
the magnification.
The extent of the diffuse emission is estimated from residual maps (Section~\ref{sec:extendedcomp}) and should be regarded as approximate.
}
\label{tab:characteristic_scales}
\end{table}

The blue component is consistently compact but mostly resolved, with effective radii of  $0\farcs03$ in the F200W band and increasing to $\sim 0\farcs05$ at longer wavelengths (F356W). 
The only exception occurs in the bluest filters (F090W/F115W), where a PSF model provides an equally acceptable fit, implying $r_{\rm e} \lesssim 0\farcs015$ (half the PSF FWHM of F090W). Toward longer wavelengths, 
the two components become increasingly blended, as the PSF broadens 
and the red component becomes brighter relative to the blue.

Constraining the size of the red component is challenging. In the 
SW bands, faint surrounding emission introduces a degeneracy 
between its intrinsic extent and the low-surface-brightness 
emission (Appendix~\ref{app:galfit_tests}). In the LW bands, 
the red component dominates the flux, but the larger PSF leads 
to poor constraints. For these reasons, we adopt a PSF parameterization for the fiducial model, while emphasizing that the current data do not uniquely distinguish between a truly unresolved source and a very compact resolved component. Nevertheless, we can place a 
conservative upper limit using F444W, the band where the red 
component is brightest and least affected by faint extended emission 
(Section~\ref{sec:extendedcomp}). Adopting half the PSF FWHM 
($\sim0\farcs14$) as an approximate resolution limit, we 
obtain $r_{\rm e} \lesssim 0\farcs07$. 


Taken together, the results reveal a structurally complex system in which the red and blue components sit in close proximity, separated by only $\sim0\farcs18$.  Their structural parameters vary with wavelength, hinting at different physical contributions to the emission in each band, such as compact stellar and/or AGN continuum, along with potential extended nebular-line emission, as we explore in 
Section~\ref{sec:sedcomponents}.

\subsection{Evidence for Low-surface-brightness Emission}
\label{sec:extendedcomp}

The current model includes two distinct components: the blue and red components. Beyond these, there is evidence for an additional faint extended structure. 
A first indication is a faint “bridge”-like structure connecting the two clumps, which is already visible in both multiple images (A and B; see Appendix~\ref{app:imageA}). 
This emission, while essentially attached to the compact red component, has a distinct color, suggesting a physically different origin of emission.

A second indication comes from the segmentation masks of the SW bands (see Figure~\ref{fig:galfit_final}), where a consistently detected region appears in the southeast direction, with an extent of at least $\sim 0\farcs3$. A similar hint of extended flux is visible in the F277W and F356W images.

This extended emission becomes clearer in the residuals, obtained by subtracting the convolved model from the science image. In Figure~\ref{fig:galfit_final} (bottom panel), we stack the residuals separately for the SW and LW filters. In both stacks, the faint structure is recovered with a similar extent (shown in green). In particular, for the SW residual stack, we recover the bridge emission between the red and blue components (overplotted as open circles for visualization).  We measure a characteristic extent of $d_{\rm extend}\sim 0\farcs3$, with emission detectable out to $\sim 0\farcs5$ (measured directly from the stacked residual maps, without PSF correction).

In particular, these features are harder to detect in A383-LRD1A due to the significantly higher background of the surrounding ICL, but it is present. Due to the higher magnification and the parity flip (see Section~\ref{sec:lensingmodel}), it appears on the opposite side and at a larger apparent separation. This provides further confirmation that the emission is physically associated with this LRD (see Appendix~\ref{app:imageA}).

To quantify this excess emission, we measure the residual flux in each band within the same aperture radius as for the total photometry ($0\farcs5$), with uncertainties estimated from the background rms. The resulting SED of the extended component is discussed in Section~\ref{sec:sedcomponents}.

Finally, independent support for an extended structure comes from ground-based spectroscopic observations. DEIMOS spectroscopy reveals spatially extended Ly$\alpha$ emission for A383-LRD1B with a diameter of $\sim 3 \arcsec$ (corresponding to $\sim 5$ kpc 
in the source plane) along the slit direction \citep{Richard2011}. Notably, the DEIMOS slit was aligned roughly along the same southeast direction as the extended feature identified in our JWST segmentation maps. Recently, \citet{Knudsen2025} reported spatially extended 
[O\,III]\,88$\mu$m emission in A383-LRD1A from 
Atacama Large Millimeter/submillimeter Array (ALMA) observations. The spatial extent of their map is consistent 
with the morphology observed in filter F356W, where the optical 
[O\,III]+H$\beta$ complex falls at this redshift and where 
the residual emission is most prominent. 

All of this suggests that the system hosts additional, low–surface-brightness emission beyond the two compact components modeled in our fits, though this structure is only weakly detected in the current JWST imaging.

\begin{figure*}
    \centering
    \includegraphics[width=0.48\linewidth]{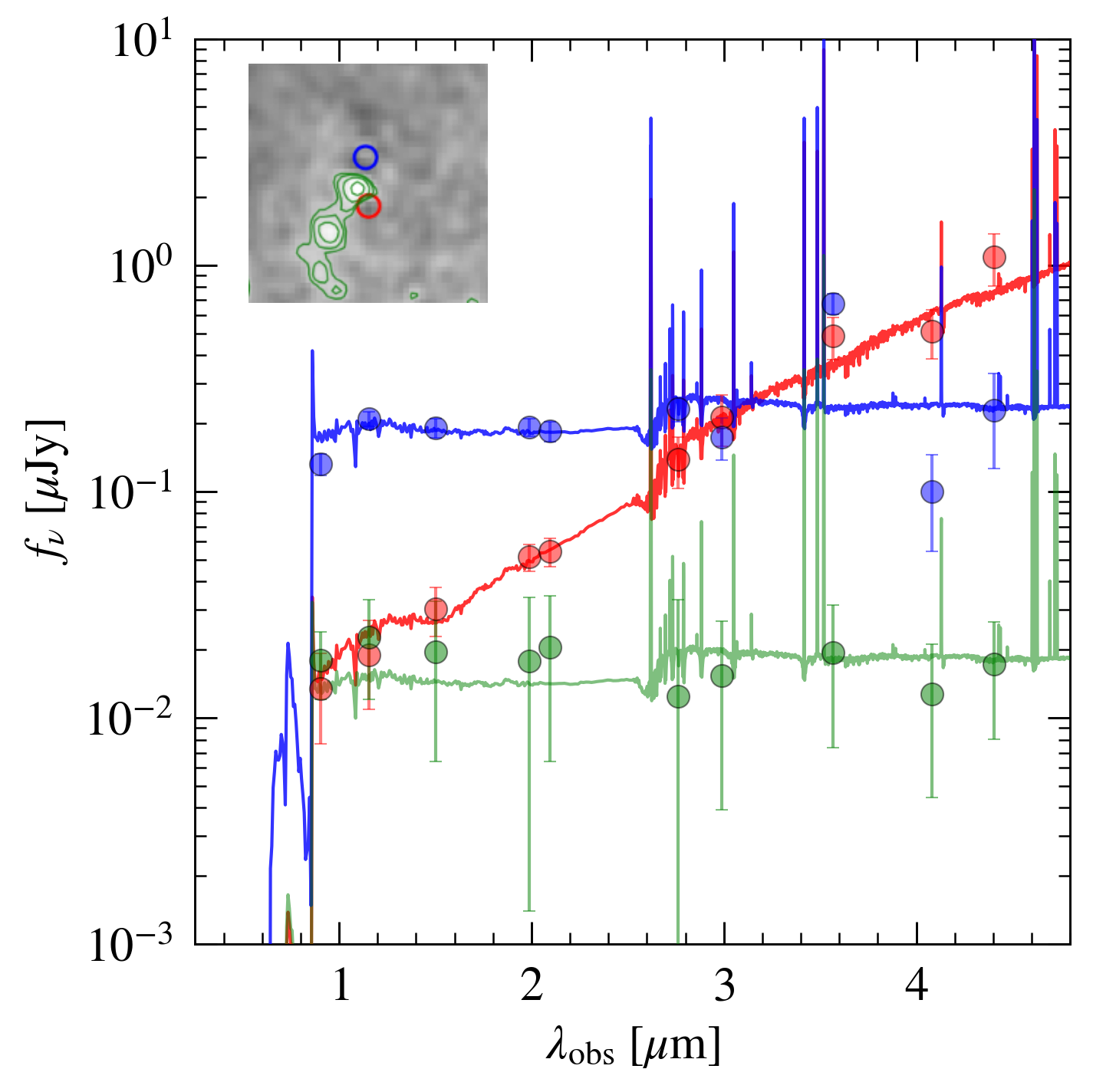}
\includegraphics[width=0.48\linewidth]{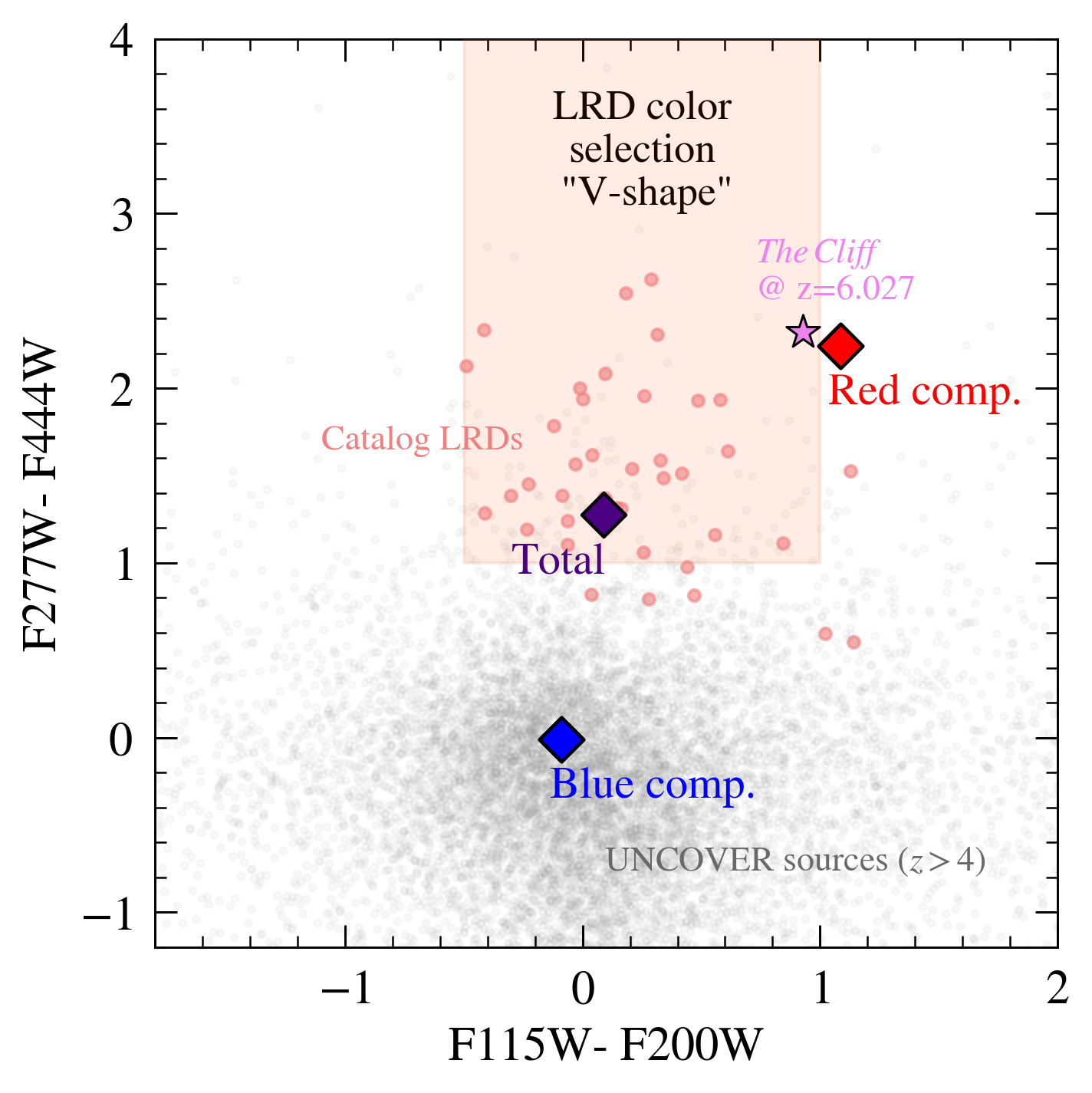}
    \caption{\textit{Left}: SED decomposition of the spatially resolved components derived from GALFIT modeling (red, blue, residual), shown as colored points. Solid lines show best-fit \eazy\, fits, which assume simple stellar population templates. The red component has a steeply rising SED, while the blue component shows a relatively flat rest-frame UV continuum, characteristic of young, unobscured star formation. \textit{Right:} color--color diagram showing $m_{\rm F115W} - m_{\rm F200W}$ vs. $m_{\rm F277W} - m_{\rm F444W}$.
  The purple diamond shows the total colors of A383-LRD1B from aperture photometry; red and blue diamonds show the       
  decomposed GALFIT components. The pink shaded box indicates the ``V-shaped'' LRD selection criterion of \citet{Greene2024}.
  Gray points show UNCOVER DR3 sources at $z_{\rm phot} >
   4$; pink circles show LRDs from \citet{Kocevski2025ApJ...986..126K} and \citet{Greene2024}. The pink star indicates synthetic NIRCam colors computed from the NIRSpec/PRISM spectrum of RUBIES-UDS-154183 \citep[``The
  Cliff'';][]{Graaff2025_Cliff}, shifted to $z = 6.027$ (see text).
    }
    \label{fig:component_seds}
\end{figure*}


\section{Spectral Energy Distribution: Splitting and Fitting}
%
\label{sec:sedcomponents}
Having established the structural decomposition into separate components, we next determine their spectral properties, focusing first on the blue and red components. To measure the light from each component across all filters, we fix the profile parameters to their F200W values and allow only the total magnitude to vary.  This differs from the full structural
fitting described in Section \ref{sec:morphparameters}, where all parameters
were left free; here we instead perform forced photometry in the remaining bands. 
Flux uncertainties are taken from the GALFIT simulations described above, which provide the 16th–84th percentile range of the recovered magnitudes for each component in each band. 
These magnitudes and uncertainties are then converted to flux densities to construct the component SEDs.

Figure~\ref{fig:component_seds} (left) shows the resulting SEDs, with individual flux measurements from the decomposed forced photometry shown as scatter points. 
To guide the interpretation of the underlying stellar populations, we fit these fluxes using \eazy\, \citep{Brammer2008}, with standard FSPS-based template sets (tweak\_fsps \_QSF \_12 \_v3), as well as with \bagpipes\, \citep{Carnall2018}.
The models assume purely stellar emission and exclude any AGN contribution. 
The best-fit models, shown in solid, are fit directly to the observed fluxes, but the physical parameters quoted below are derived from the intrinsic (delensed, assuming $\mu_B=7.3$) SEDs.

At short wavelengths, the emission is dominated by the blue component, which shows a nearly flat SED in $f_\nu$. It shows an enhancement in F356W, likely due to the H$\beta$+[O\,III] complex, and remains relatively flat toward the red. The blue component is well fit by a young, star-forming stellar population with low dust attenuation ($A_V \sim 0.07$), a stellar mass of $\mstar \sim 3\times10^8 \msun$, and a star formation rate of $\mathrm{SFR}\sim1~\msun~\mathrm{yr}^{-1}$,
and we measure a UV continuum slope of $\beta_{\rm UV} \approx -2$, typical of high-redshift star-forming galaxies and clumps \citep[e.g.,][]{Bolamperti202310.1093/mnras/stad3114,Cullen2023MNRAS.520...14C}. 
In contrast, the red component becomes increasingly dominant toward longer wavelengths with an extremely steep SED, overtaking the blue around the Balmer break. 
The best-fit model suggests a significantly older, more massive ($ \mstar \sim 2\times 10^{10}\msun$), and more dust-attenuated population ($A_V\sim 2.6$).


Based on the HST+JWST photometry alone, the SED admits a straightforward stellar interpretation, with a blue, unobscured component and a red component that is massive and heavily dust attenuated.
However, this picture becomes challenged when incorporating submillimeter constraints.
A stellar population with this level of dust attenuation would be expected to produce substantial far-IR emission, which is not observed given the ALMA continuum upper limit \citep{Knudsen2016_10.1093/mnrasl/slw114}. Enforcing this constraint leads to poorer fits to the optical part of the SED for the red component and reduces the inferred stellar mass by more than an order of magnitude. We discuss this in more detail in Appendix~\ref{app:sed_fitting_bagpipes_alma}.

The extremely red SED may therefore also point to a nonstellar origin such as a reddened AGN.
The red colors and steep Balmer breaks have recently been interpreted as signatures of accreting black holes embedded in dense gas configurations \citep{Juodzbalis2024,Graaff2025_Cliff,Graaff2025_unified,InayoshiMaiolino2025,Ji2025,Kido2025MNRAS.544.3407K, Liu2025ApJ...994..113L, Naidu2025_BHstar,
 Asada2026arXiv260110573A,
 DEugenio2025,
 Inayoshi2026ApJ..1000...90I, 
 Kokorev2025arXiv251107515K,
 Liu2026arXiv260302317L,
 MadauMaiolino2026arXiv260222386M, 
 Matthee2026arXiv260317667M,
 Pacucci2026arXiv260114368P,
 Rusakov2026Natur.649..574R,
Sneppen2026arXiv260118864S,  Sun2026arXiv260120929S}.

It is intriguing that this component begins to dominate exactly at the Balmer break, the same wavelength at which the characteristic inflection appears in LRD SEDs \citep{Setton2025ApJ...995..118S_vshape}. If the emission is a stars+AGN composite, it is not obvious why the AGN would begin to dominate precisely at this point, though this recurring coincidence across LRDs may hint at a more fundamental connection.

We now explore the SED of the faint component (as discussed in Section~\ref{sec:extendedcomp}). We use residual fluxes derived from the fixed two-component model, ensuring a consistent decomposition across all bands. The resulting SED is relatively flat in $f_\nu$, similar to the blue component, and contributes a small but measurable fraction of the total light ($\sim$4\% on average, reaching up to $\sim$7\% in some bands). Notably, in F090W the residual component is comparable in 
brightness to the red component, highlighting the degeneracy 
in flux assignment between the two. A stellar fit gives $ \mstar \sim 2\times 10^{7}\msun$, though we caution that the per-band signal-to-noise ratio is low, making a robust physical interpretation uncertain.

Figure~\ref{fig:component_seds} (right) presents a color--color diagram. The box indicates the ``V-shaped'' LRD selection criterion of \citet{Greene2024}, defined by $-0.5 < m_{\rm F115W} - m_{\rm F200W} < 1.0$   
 and $m_{\rm F277W} - m_{\rm F444W} > 1.0$.
 Gray points show all UNCOVER Data Release 3 (DR3) sources \citep{Bezanson_uncover2024, Suess2024_mediumbands, Weaver2024} with  $z_{\rm phot} > 4$. The total colors for A383-LRD1B, derived from aperture photometry, are shown as a purple diamond, while the decomposed red and blue components from the forced GALFIT  photometry are shown as red and blue diamonds, respectively.        The red component falls just outside the selection box, while the blue component coincides with the locus of the high-redshift UNCOVER sources.   For comparison, we show synthetic colors computed from the NIRSpec/PRISM spectrum of the LRD RUBIES-UDS-154183 at $z = 3.55$ \citep[``The Cliff'';][]{Graaff2025_Cliff}. To place it on the same diagram, we shift the observed wavelength axis to $z =
  6.027$ and compute filter-weighted synthetic fluxes, which is equivalent to asking what colors this source would have if observed at the redshift of A383-LRD1. The resulting colors fall very close to those of the red component, supporting a similar intrinsic spectral shape.
  Finally, we show in pink circles selected LRDs by two independent studies \citet{Kocevski2025ApJ...986..126K, Greene2024}, which cluster around the total colors of A383-LRD1.

Fully disentangling these components and determining the origin of their emission will require spatially resolved spectroscopy. 
The SEDs shown here should therefore be taken as indicative rather than definitive. Nevertheless, the relative SED shapes and colors already prove to be highly informative. The blue component remains remarkably stable across the modeling (see Appendix~\ref{app:galfit_tests}), demonstrating that its inferred SED shape is robust and largely insensitive to assumptions about the decomposition. 
Most importantly, the SED decomposition suggests that the iconic “double-break’’ (or “V-shaped’’) SED (in $f_{\lambda}$) of this LRD is not an intrinsic feature of a single component. Instead, it arises from the combined light of two physically distinct sources.

\begin{figure*}
    \centering
    \includegraphics[width=\linewidth, trim={7cm 0cm 3cm 0cm},clip]{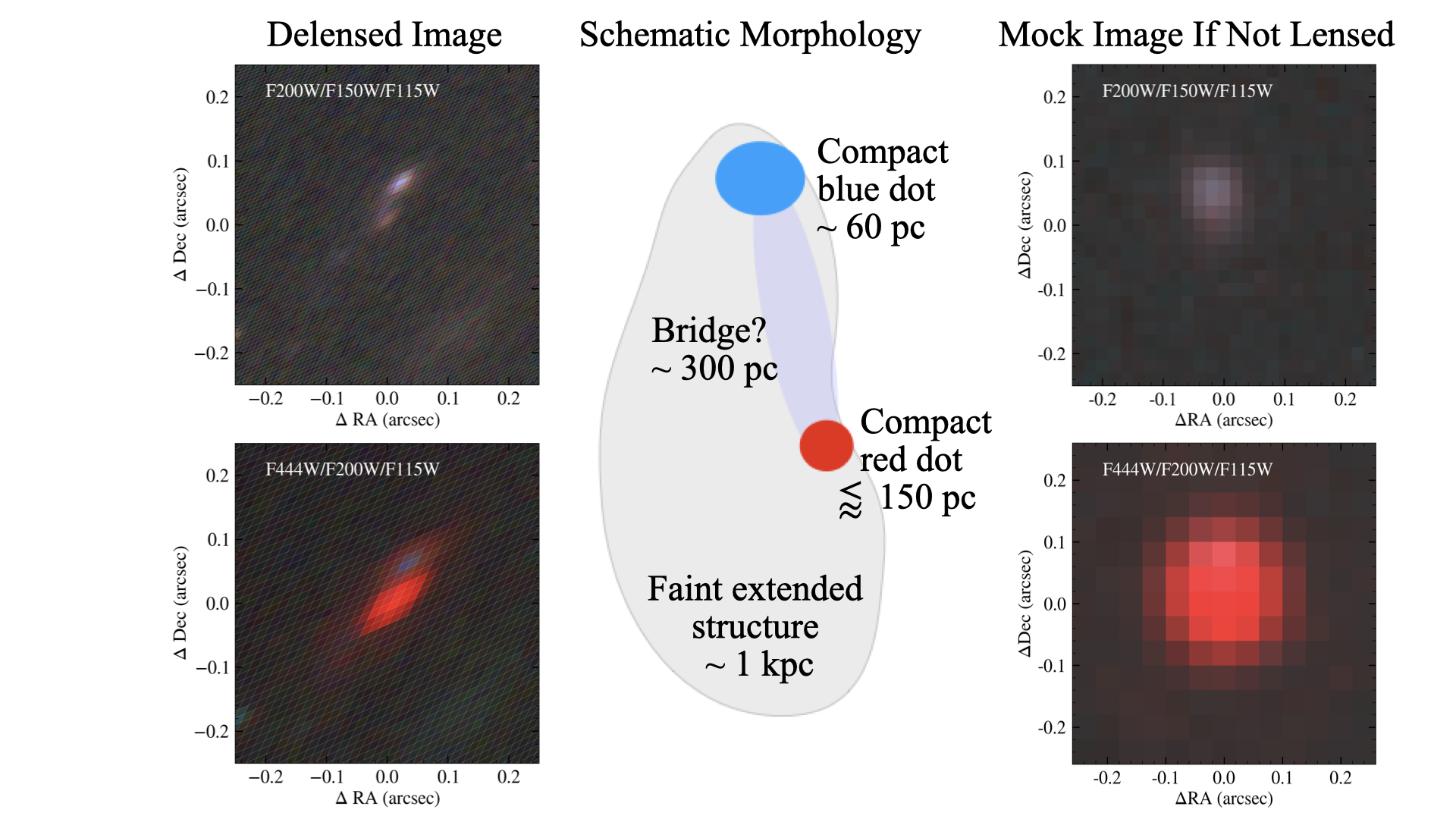}
    \caption{\textit{Left}: delensed warped-pixel reconstruction of the observed RGB images (F200W/F150W/F115W, top; F444W/F200W/F115W, bottom). Each pixel is mapped back to the source plane via the lens equation.
    \textit{Middle}: schematic reconstruction of the intrinsic morphology inferred from multiband GALFIT modeling combined with lensing corrections.
    \textit{Right}: mock RGB images (F200W/F150W/F115W, top; F444W/F200W/F115W, bottom) in the absence of lensing. The best-fit models for each filter are corrected for lensing (see text), convolved with the regular JWST PSF, and realistic background is added. These images demonstrate that, in the absence of magnification, all substructure would be compressed to scales comparable to the native JWST pixel size, making the intrinsic morphology challenging to uncover.
    }
    \label{fig:warped_pix_intrinsic_morphology}
\end{figure*}

\section{source-plane reconstruction}
\label{sec:lensmodeling_reconstruction}

\subsection{Lens Model}
\label{sec:lensingmodel}

We now turn to the intrinsic, delensed properties of the LRD.
For this analysis, we use the publicly available A383 lensing models released through the CLASH program \citep{Zitrin2015}.\footnote{https://archive.stsci.edu/missions/hlsp/clash/a383/models/} 
We adopt the average magnification from the two parametric reconstructions provided: \texttt{Lenstool} \citep{Jullo2007_lenstool} and \texttt{GLEE} \citep{SuyuHalkola2010_GLEE}.
The lens models provide maps of total magnification ($\mu_{\rm tot}$), convergence ($\kappa$), shear ($\gamma$), and deflection fields ($\alpha_{\rm x},\alpha_{\rm y}$), computed at a source redshift of $z=2.55$. 
Since our source is at $z=6.027$ rather than the model redshift of $z=2.55$, we apply a redshift-dependent scaling to the deflection, shear, and convergence maps, given by the ratio of angular diameter distances:
\begin{equation}
   \beta = \frac{d_{\rm A}(z_{\rm lens}, z_{\rm s})/d_{\rm A}(0, z_{\rm s})}{d_{\rm A}(z_{\rm lens}, z_{\rm model})/d_{\rm A}(0, z_{\rm model})}.
\end{equation}
 Here $d_{\rm A}$ is the angular diameter distance between two redshifts, and with  $z_{\rm lens}=0.189$ (A383), 
 $z_{\rm model}=2.55$, 
 and our source at $z_{\rm s}=6.027$, we find $\beta=1.04$.

After applying this correction (e.g., $\kappa \rightarrow \beta \kappa$, $\gamma \rightarrow \beta \gamma$), we compute the tangential magnification as $\mu_{\rm tan} = 1/(1-\kappa -\gamma)$ and radial magnification as $\mu_{\rm rad} = 1/(1-\kappa +\gamma)$.
The total magnification is given by the product $\mu_{\rm tot } = \mu_{\rm tan} \mu_{\rm rad} = 1/((1-\kappa)^2-\gamma^2)$. 
Gravitational lensing magnifies the galaxy in both area and flux by $\mu_{\rm tot }$ while conserving surface brightness, such that $m_{\rm obs} = m_{\rm int} - 2.5 \log (\mu_{\rm tot })$.

For A383-LRD1A, we find $\mu_{A, \rm tan} = -4.0$ (negative indicating parity flip), $\mu_{A, \rm rad} = 2.8$, and  $|\mu_{A,\rm tot}| = 10.8$. 
For A383-LRD1B, we find $\mu_{B,\rm tan} = 4.3$, $\mu_{B,\rm rad} = 1.7$, and  $\mu_{B,\rm tot} = 7.3$. The total magnification values are in agreement with the reported values by \citet{Richard2011}. In addition, the  predicted magnification ratio $\mu_{A,\rm tot}/\mu_{B,\rm tot}$ is also consistent with the observed A383-LRD1A/A383-LRD1B flux ratio, which is stable across all bands at $\sim$1.4--1.6, providing confidence in the overall normalization of the model magnifications.
We note, however, that substantial variations exist among different lens modeling methods, both in the total magnification as well as in the decomposition into radial and tangential components. 
Typical uncertainties in cluster-core magnifications are of order $\sim20\%$ \citep{Zitrin2015}, representing an additional systematic uncertainty on all intrinsic quantities reported here.


\subsection{Direct Ray-tracing of the JWST Pixels}

To visualize the morphology implied by the lens model with minimal assumptions, we perform a direct, pixel–based ray-tracing reconstruction of the JWST images. For each detector pixel, we compute the source-plane coordinate via $\beta(\theta) = \theta - \alpha(\theta)$ and trace its four pixel corners using the deflection maps ($\alpha_x,\alpha_y$). This maps every image-plane pixel to an irregular quadrilateral in the source plane. The procedure removes only the geometric distortions introduced by lensing while conserving surface brightness and retaining the effects of the JWST PSF and pixel sampling.

Figure~\ref{fig:warped_pix_intrinsic_morphology} (left panel) shows the resulting warped-pixel 
reconstruction for  
a rest-frame UV image  (F200W/F150W/F115W), analogous to Figure~\ref{fig:morphology}, and for a redder-color composite image (F444W/F200W/F115W).
These examples use the \texttt{Lenstool} model, and we confirm that the \texttt{GLEE} model yields qualitatively consistent results.
In the UV warped-pixel reconstruction, the two dominant components remain clearly detected and offset from one another, consistent with the image plane. 
In the optical combination, the redder emission becomes more prominent.
While this warped-pixel representation provides an intuitive visualization of the lensing geometry, it is not used for quantitative measurements of the intrinsic structure.
The intrinsic morphology is instead derived by fitting the components in the observed plane with PSF-corrected GALFIT models and then applying the magnification corrections, as we will show in the next section.

\subsection{Intrinsic Sizes and Separations from Magnification Corrections}
\label{sec:intrinsicproperties}
We now apply magnification corrections to the observed sizes and separations derived from the best-fit GALFIT models (Section~\ref{sec:morphparameters}). These are then converted into physical scales assuming a scale of $1\arcsec = 5.8$ kpc at the source redshift of $z = 6.027$.
The resulting intrinsic values are summarized in Table~\ref{tab:characteristic_scales}.

To compute the intrinsic separation between the red and blue components, we account for directional magnification along the measurement axis. The observed separation is corrected using $\mu_{\theta}=\sqrt{(\mu_{\rm tan}\cos(\theta))^2 + (\mu_{\rm rad}\sin(\theta))^2}$, where $\theta$ is the angle between the component separation vector and the local tangential direction. This gives a directional magnification $\mu_\theta$ between $\mu_{\rm tan}$ (purely tangential, $\theta=0^\circ$) and $\mu_{\rm rad}$ (purely radial,  $\theta=90^\circ$). The intrinsic separation is then given by $d_{\rm int} = d_{\rm obs} / \mu_{\theta}$.
We estimate $\theta$ by determining the local shear orientation from the deflection maps. At the position of the red component in A383-LRD1B, we find $\theta = 33^\circ$, resulting in $\mu_\theta = 3.7$.
For A383-LRD1A, we find $\theta = 73^\circ$, corresponding to $\mu_\theta = 2.9$.

Applying these corrections, we find an intrinsic separation of $\sim0\farcs05$ for A383-LRD1B, or about $\lesssim$3 SW pixels. For A383-LRD1A, the inferred separation is twice as large. Because A383-LRD1A lies very close to the critical curve, its local magnification is likely less reliable than for A383-LRD1B. Moreover, A383 is known to produce prominent radial arcs \citep{Sand2005, Sand2008}, and A383-LRD1A falls in a region where radial stretching may be significant and potentially underestimated. We therefore adopt the A383-LRD1B measurements as the more reliable estimate of the intrinsic separation. Further, the scale of $\sim$300 pc is identical to the total half-light radius previously reported by \citet{Richard2011} using HST imaging.

For the intrinsic sizes of the blue and red components, the position angles are poorly constrained due to their compact and nearly circular morphologies. We therefore adopt an isotropic magnification correction, dividing the observed sizes by $\sqrt{\mu_{\rm tot}}$. 

After correction, both components are found to be physically extremely compact. In the rest-UV, 
the blue component has a characteristic size of $r_{\rm e,circ}\sim60$ pc, ranging from $\sim50$ to $70$ pc across the SW bands.
Upper-limit constraints for the red component are obtained from the F444W filter, yielding $r_{\rm e,circ}\lesssim 150$ pc. We interpret this as the most robust upper limit on the intrinsic compactness of the red component, since the surrounding low-surface-brightness emission contributes relatively the least to the total light in this band. The extent of this low-surface-brightness emission, in which the compact knots are embedded, reaches roughly 1 kpc in the source plane.

A schematic representation of the inferred source-plane morphology is shown in Figure~\ref{fig:warped_pix_intrinsic_morphology} (middle panel). The system consists of two ultracompact dots,  connected by a $\sim$300 pc bridgelike component and embedded within a more diffuse structure that extends to scales of up to $\sim$1 kpc. This interpretation is based on the available imaging data; the detailed spatial distribution of individual emission lines remains to be mapped with future spectroscopy.

\section{Discussion}
\label{sec:discussion}
The central result of this paper is the discovery of a newly identified LRD that is doubly lensed by the galaxy cluster A383. The two images, A383-LRD1A and A383-LRD1B, are predicted to have total magnifications of $\mu_A \sim 11$ and $\mu_B \sim 7$, respectively, making this source one of the most highly magnified LRDs known. 

Thanks to this extreme magnification, the system does not appear as a single “little red dot’’ but instead breaks into two compact components: a red dot and a spatially offset blue dot, separated by only $\sim300$ pc in the source plane. Both components are extraordinarily small. In the rest-UV, the effective radius of the blue component is only a few tens of parsecs, 
 similar to the upper-limit constraints of $\sim$35~pc \citet{Furtak2023_triple}, and
comparable to the sizes of high-redshift star clusters and clumps \citep[e.g.,][]{Claeyssens2023,Vanzella2023}. The red component is also highly compact, with a rest-optical upper limit of $r_{\rm e}<150$~pc from the F444W imaging.

Turning to the origin of the emission, the physical nature of these components remains under investigation. The blue component likely corresponds to a compact star cluster. This is supported by its consistent morphology across all UV bands and a flat SED, indicative of continuum-dominated emission from a young stellar population. 

The red component, on the other hand, is more enigmatic. One interpretation, motivated by its extreme compactness, is that it may host an accreting black hole. This scenario could be tested directly via high-resolution spectroscopy capable of resolving Keplerian rotation, as recently demonstrated in \citet{Juodvzbalis2025_BH_measurement}. Its SED is indeed very steep, resembling that of other compact red sources such as the The Cliff/BH* systems \citep{Graaff2025_Cliff, Naidu2025_BHstar}, and may be indicative of a dense gas environment \citep[e.g.][]{InayoshiMaiolino2025, Ji2025, Kido2025MNRAS.544.3407K, Liu2025ApJ...994..113L,Rusakov2026Natur.649..574R, Sneppen2026arXiv260118864S}. 
Remarkably, the red component resides in an environment characterized by low metallicity, low dust content, and a strong UV radiation field implied by the high [O\,III]/[C\,II] ratio \citep{Knudsen2016_10.1093/mnrasl/slw114,Knudsen2025}. The blue component, just
  300~pc away, is the most natural source of this radiation. In \citet{Baggen_2026}, the Lyman--Werner flux is quantified ($\log J_{21} = 4.4$) and linked to conditions consistent with direct-collapse black hole formation across a large sample of LRDs including this system.


Alternatively, we explore a stellar-only interpretation for the red component.
The presence of a Balmer break in the SED suggests that the F277W morphology may trace the older stellar component. The best-fit stellar mass of $\mstar \sim2 \times 10^{10}~M_\odot$ within an effective radius ($r_{\rm e}\leq150\rm \, pc$) would imply high stellar densities ($\Sigma_{\rm eff}\sim 2\times10^5 \msun\rm pc^{-2}$) and therefore require substantial dynamical support. Using the scaling arguments of \citet{Baggen2024}, such a system would be expected to 
have FWHM line widths of $\geq$ 1000 km s$^{-1}$ and be resolvable across $\sim$150 pc scales. However, if the stellar mass is enclosed within a smaller radius, the required velocity dispersions would be even higher.

A third possibility is that the optical emission arises from a composite system, consisting of a massive stellar cluster hosting an actively accreting black hole. Such a configuration is not only plausible but is in fact predicted by recent theoretical models, in which dense stellar systems assemble efficiently in a feedback-suppressed regime and then provide favorable conditions for early black hole growth \citep{Dekel2023, Dekel2025, Pacucci2025}.
 The morphology observed here, with its multi-component structure and, in particular, the size and stellar mass of the blue component, is in striking agreement with the characteristic properties predicted by these models.

Altogether, these results motivate a straightforward observational question: \textit{What would this system look like in the absence of strong lensing?}
Figure~\ref{fig:warped_pix_intrinsic_morphology} (right panel) illustrates this. Here we take the best-fit intrinsic model by applying the appropriate magnification corrections to the observed sizes and fluxes. 
We map the result onto a regular grid, convolve it with the JWST PSF, and add realistic background noise. We then generate an RGB composite image using the F200W/F150W/F115W and F444W/F200W/F115W filter combinations. In the rest-UV composite (top panel), the red component, already faint in the rest-UV in the models and further suppressed after demagnification, becomes effectively invisible, leaving the apparent structure dominated by the blue UV clump. In the LW composite (bottom panel), the system reduces to a single PSF-like blob. In both cases, the morphology collapses into a nearly pointlike, blended source, such that separating the red and blue components would be extremely difficult.


This has important implications for the broader LRD population. Features that we can spatially resolve here would normally be blended together. The most important example is the characteristic V-shaped SED. In this system, the V-shape is spatially resolved: Our component-based SED decomposition shows that it is simply the superposition of the blue and red components \citep[see also][who propose a similar interpretation]{Barger2026arXiv260527903B,Chisholm2026arXiv260215935C, Sun2026arXiv260120929S}.
The same holds for the fluxes in F277W and F356W, filters that may trace the Balmer break and emission-line complexes such as [O\,III]+H$\beta$, respectively, where both components contribute comparable flux.
As a result, emission-line properties commonly attributed to a single compact region in LRDs need not originate from a single physical component. Instead, we might even expect the observed spectra to reflect a mixture of line widths, equivalent widths, and apparent Balmer-break signatures arising from physically distinct regions, rather than requiring one region to simultaneously reproduce all characteristic features.


Looking ahead, this system provides an important benchmark for interpreting LRDs more generally. Improved lens modeling for A383, by incorporating the new JWST imaging, will further refine the magnification field, and hence the intrinsic sizes and separations. High-resolution spectroscopy, ideally with JWST/NIRSpec IFU, will be crucial for disentangling the stellar and nonstellar contributions to the emission, mapping their kinematics, and linking the spatially resolved structure to the integrated emission-line diagnostics used for unlensed LRDs. More broadly, the discovery of additional strongly lensed LRDs with resolved internal structure \citep[see also][]{Baggen_2026,Barger2026arXiv260527903B, Yanagisawa2026}
will be essential to determine whether the configuration seen here is a rare phenomenon or a common configuration of LRDs.


\begin{acknowledgments}
\software{Astropy \citep{astropy:2013, astropy:2018, astropy:2022}, 
Matplotlib \citep{Matplotlib_Hunter2007}, SciPy \citep{SciPy_Virtanen2020}, 
NumPy \citep{2020NumPy-Array}, 
colossus \citep{Diemer2018_colossus} and photutils \citep{larry_bradley_2025_14889440}}.
This work is based (in part) on observations made with the NASA/ESA/CSA James Webb Space Telescope. The data were obtained from the Mikulski Archive for Space Telescopes at the Space Telescope Science Institute, which is operated by the Association of Universities for Research in Astronomy, Inc., under NASA contract NAS 5-03127 for JWST. These observations are associated with program GO 6882. 
The specific observations can be accessed via \dataset[doi:10.17909/5a2a-dp32]{https://doi.org/10.17909/5a2a-dp32}.
\end{acknowledgments}

%



\appendix

\section{GALFIT modeling tests}
\label{app:galfit_tests}

 Our primary goal is to characterize the morphology and relative flux contributions of the red and blue components.    
Decomposing this system requires several decisions: the functional form for each component (S\'ersic or PSF), the PSF model, and the number of components included. 
Each choice introduces systematic uncertainties that in several bands exceed the statistical uncertainties from injection-recovery simulations. While the absolute fluxes and sizes of individual components are therefore not uniquely determined, the presence of two spatially distinct components with clearly different colors is robust across all models considered.

\begin{figure*}[h!]
      \centering
      \begin{minipage}{0.48\textwidth}
          \centering
          \includegraphics[width=\linewidth]{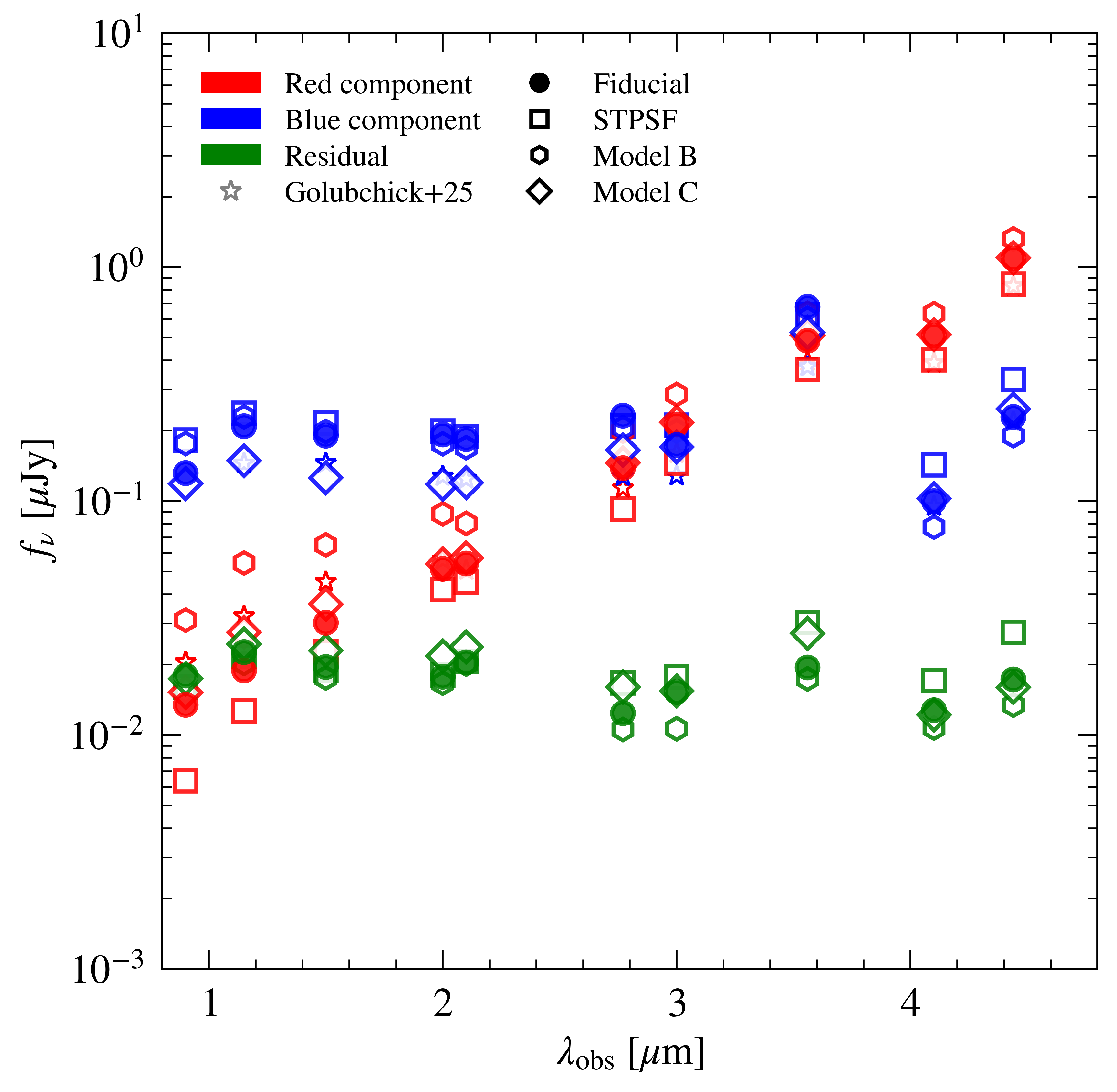}
          \caption{Component SEDs derived from different GALFIT model configurations discussed in this appendix.}
          \label{fig:app:seds}
      \end{minipage}
      \hfill
      \begin{minipage}{0.5\textwidth}                     \centering
    \begin{tabular}{cccc}
   \hline
    Band & Red & Blue & Residual \\
    \hline
    F090W & $28.6 \pm 0.5 \pm 0.7$ & $26.1 \pm 0.1 \pm 0.2$ & $28.3 \pm 0.4 \pm 0.1$ \\
    F115W & $28.2 \pm 0.5 \pm 0.9$ & $25.6 \pm 0.1 \pm 0.2$ & $28.0 \pm 0.5 \pm 0.1$ \\
    F150W & $27.7 \pm 0.3 \pm 0.6$ & $25.7 \pm 0.1 \pm 0.2$ & $28.2 \pm 0.7 \pm 0.1$ \\
    F200W & $27.1 \pm 0.1 \pm 0.4$ & $25.7 \pm 0.1 \pm 0.2$ & $28.3 \pm 1.0 \pm 0.1$ \\
    F210M & $27.1 \pm 0.1 \pm 0.3$ & $25.7 \pm 0.1 \pm 0.2$ & $28.1 \pm 0.7 \pm 0.1$ \\
    F277W & $26.0 \pm 0.3 \pm 0.3$ & $25.5 \pm 0.1 \pm 0.1$ & $28.7 \pm 1.8 \pm 0.2$ \\
    F300M & $25.6 \pm 0.3 \pm 0.3$ & $25.8 \pm 0.2 \pm 0.1$ & $28.4 \pm 0.8 \pm 0.2$ \\
    F356W & $24.7 \pm 0.2 \pm 0.2$ & $24.3 \pm 0.1 \pm 0.1$ & $28.2 \pm 0.7 \pm 0.3$ \\
    F410M & $24.6 \pm 0.3 \pm 0.2$ & $26.4 \pm 0.5 \pm 0.3$ & $28.6 \pm 0.7 \pm 0.2$ \\
    F444W & $23.8 \pm 0.3 \pm 0.2$ & $25.5 \pm 0.5 \pm 0.2$ & $28.3 \pm 0.6 \pm 0.3$ \\
    \hline
        \end{tabular}
\captionof{table}{Component photometry from forced GALFIT modeling. Central values are from the fiducial model (blue = S\'ersic, red
   = PSF, smoothed STPSF).
   The first uncertainty, $\sigma_{\rm rand}$ reflects the
  16th--84th percentile range from injection-recovery simulations and is used throughout the main text; the second, $\sigma_{\rm sys}$ reflects systematic variations
  across model variants discussed in this section.}
          \label{tab:appendix_magnitudes}
      \end{minipage}                                                                                      \end{figure*}

Figure~\ref{fig:app:seds} shows the component SEDs for four models: our fiducial model (blue = S\'ersic, red = PSF) fit with
  the smoothed STPSF; a S\'ersic+S\'ersic model (Model~B); a PSF+PSF model (Model~C); and the fiducial setup fit with the native STPSF. We quantify the systematic uncertainty in each band as the standard deviation of the flux across these four variants, reported alongside the recovery errors in Table~\ref{tab:appendix_magnitudes}.
  The flux of the blue component is stable across all models to within $0.2$~mag ($0.3$~mag in F410M). The red component is similarly stable in the LW bands, but shows considerable scatter in the SW bands, where bridgelike emission introduces strong degeneracies.  In the model B (S\'ersic+S\'ersic) in particular, the red component absorbs this emission, producing an artificially
  elevated flux that is likely unphysical given the distinct color of the bridge, resulting in systematic variations reaching
  $0.7$~mag in F090W and $0.9$~mag in F115W. 
    Whether the red component contributes any flux in these bands, or whether
   this emission originates entirely from the bridge, remains uncertain. 
   We also overplot aperture fluxes from \citet{Golubchik2025},
  finding good overall consistency with our decomposition.

This decomposition is made significantly easier by gravitational lensing, which magnifies the system and spatially separates the two components. This system is one of the most highly magnified LRDs known, and even here some bands remain degenerate. For unlensed LRDs, the two components would be blended (Figure~\ref{fig:warped_pix_intrinsic_morphology}), making such a decomposition considerably more challenging.

\section{Recovery of low-surface-brightness emission in image A}
\label{app:imageA}

Here we present the results for image A using the same fiducial modeling procedure adopted for image B. 
Despite the significantly higher ICL background surrounding image A, the residual morphology is consistent with that seen in image B, including extended low-surface-brightness emission and a bridgelike structure between with the compact blue and red knots. As expected from the lensing geometry, the extended emission appears on the opposite side due to the parity flip and at a larger apparent separation.
\begin{figure*}[htp!]
    \centering
    \includegraphics[width=\linewidth]{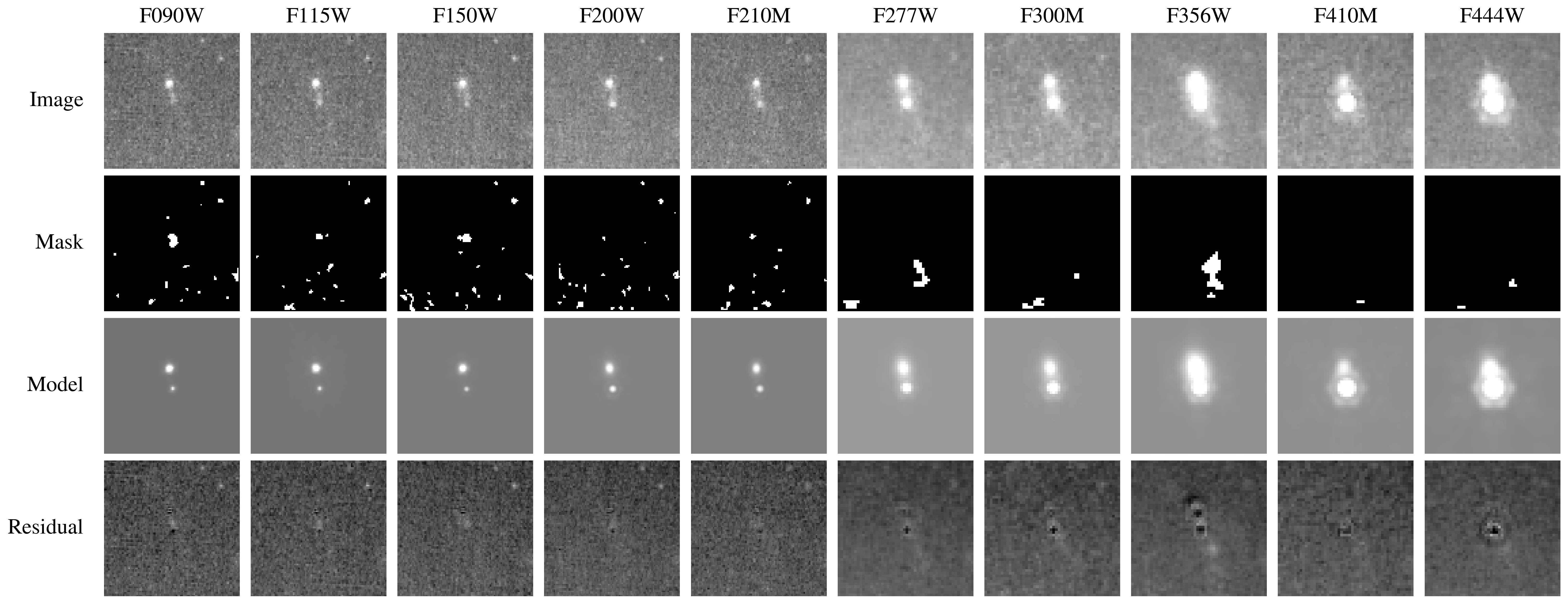}
    \includegraphics[width=0.8\linewidth]{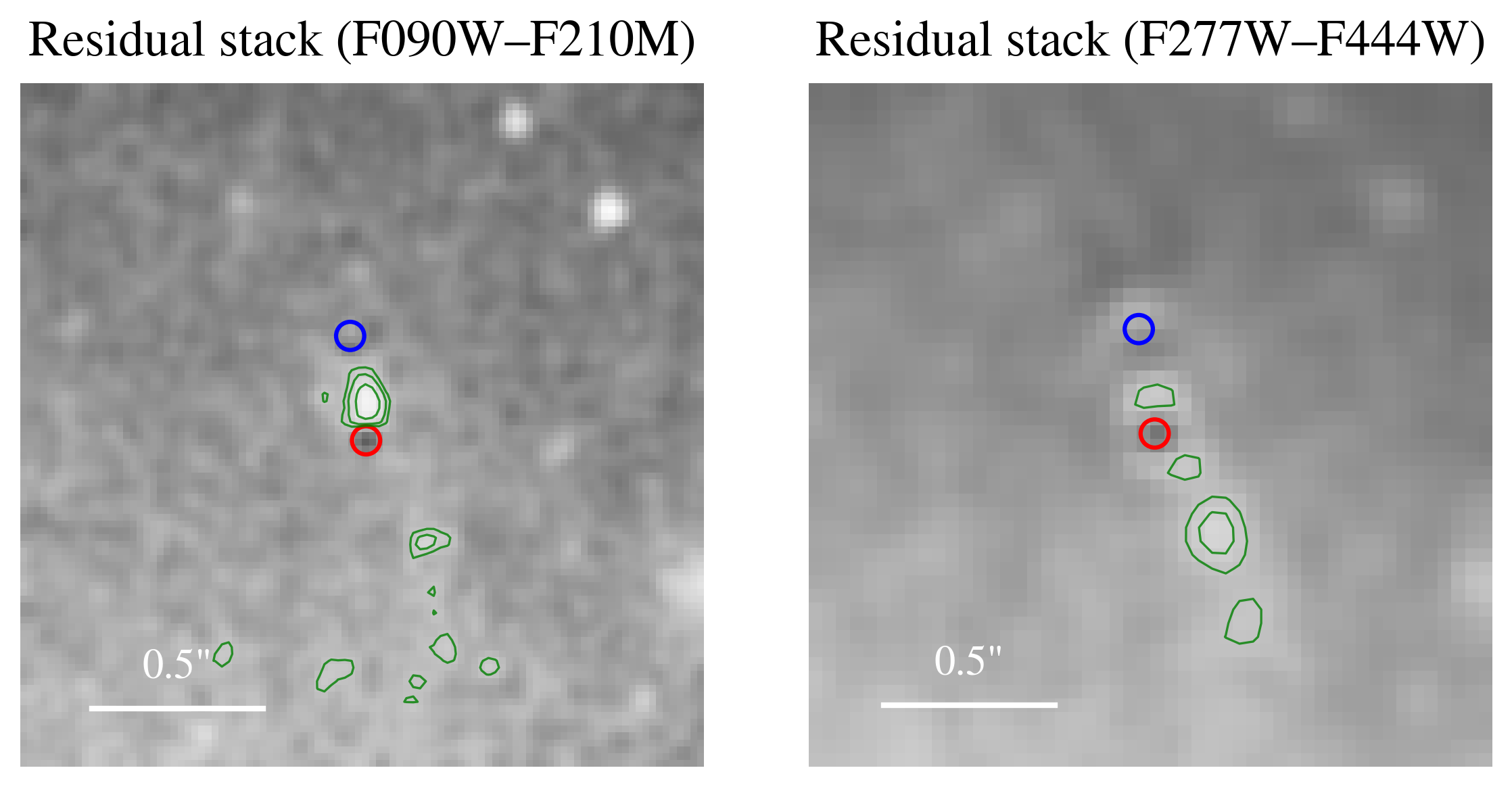}
    \caption{Same as Figure~\ref{fig:galfit_final}, but for image A.
    }
    \label{fig:app:galfit_final_IMA}
\end{figure*}

\section{SED fitting with ALMA constraints}
\label{app:sed_fitting_bagpipes_alma}

\begin{figure}
    \centering
    \includegraphics[width=0.7\linewidth]{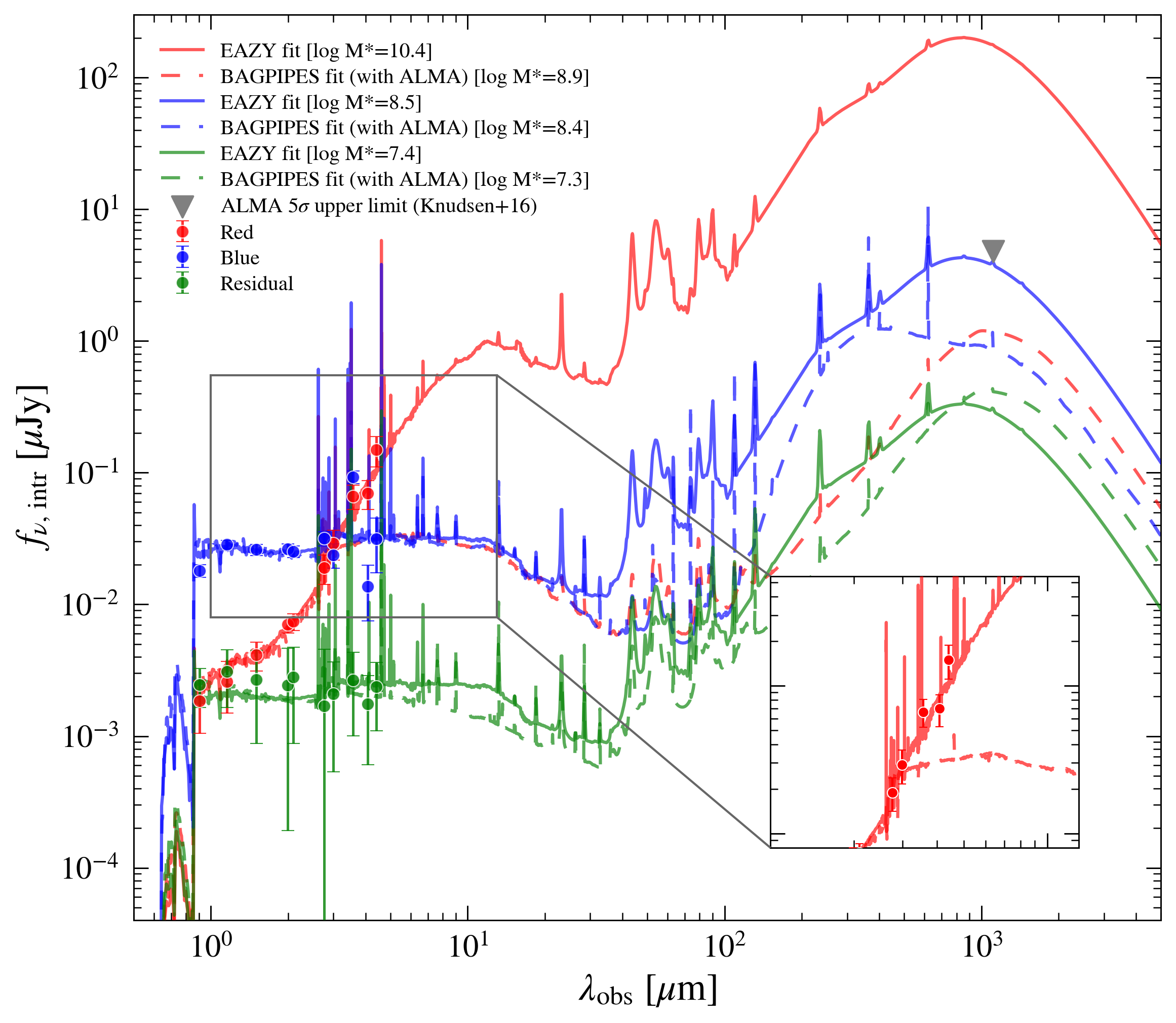}
    \caption{Spectral energy distributions of the decomposed red, blue, and residual components. Points show the measured JWST photometry. Solid curves show the \textsc{EAZY} fits, while dashed curves show the \textsc{Bagpipes} fits including the ALMA upper limit from \citet[][gray triangle]{Knudsen2016_10.1093/mnrasl/slw114}. The blue and residual components are largely unaffected by the ALMA constraint, whereas the red component is driven toward lower stellar masses while failing to reproduce the reddest JWST fluxes, illustrating the tension between the JWST photometry and the ALMA nondetection under standard stellar population models.}
    \label{fig:seds_eazy_bagpipes}
\end{figure}

We explore the impact of including the ALMA $5\sigma$ upper limit (270.445 GHz; $<55\mu$Jy) from \citet{Knudsen2016_10.1093/mnrasl/slw114}
in the SED fits in Figure \ref{fig:seds_eazy_bagpipes}. We show the \eazy\, fit, which does not include the ALMA constraint, and the \bagpipes\, fit, which includes the ALMA upper limit.
For the blue and residual components, this constraint has little effect on the inferred SEDs and stellar masses. The blue component has $\log(M_*/\msun) \approx 8.4$ in the \bagpipes\ fit and $\log(M_*/\msun) \approx 8.5$ in \eazy, while the residual component has $\log(M_*/\msun) \approx 7.3$ in \bagpipes\ and $\log(M_*/\msun) \approx 7.4$ in \eazy. In both cases, the stellar masses are consistent between the two codes within the uncertainties.
The red component, without the ALMA constraint, prefers a massive, dusty solution with $\log(M_*/\msun) \approx10.4^{+0.43}_{-0.39}$ and $A_V =2.6^{+0.27}_{-0.23}$ (\eazy\, fit), 
while including the ALMA upper limit favors
a much lower-mass, nearly dust-free solution with $\log(M_*/\msun) \approx8.9$ and $A_V \approx0.1$ (\bagpipes\, fit). 
This solution fails to reproduce the fluxes in the reddest JWST bands, as highlighted in the inset panel. This illustrates the difficulty of simultaneously matching the JWST photometry and the ALMA nondetection with standard stellar population models, a tension that has also been reported for stacked ALMA nondetections of LRDs \citep[e.g.,][]{Casey2025ApJ...990L..61C, Chen2025ApJ...994L..42C}.




    


    


\clearpage
\bibliography{sample701}{}
\bibliographystyle{aasjournalv7}



\end{document}